\documentclass[usenatbib]{mn2e}

\input epsf

\usepackage{graphicx}
\usepackage{txfonts}
\usepackage{epsfig}
\usepackage[T1]{fontenc}
\usepackage{aecompl}

\title[A test of the $\nu_{\rm max}$ scaling relation]{A test of the
  asteroseismic $\nu_{\rm max}$ scaling relation for solar-like
  oscillations in main-sequence and sub-giant stars}

\author[Coelho et al.]{H.~R.~Coelho$^{1,2}$\thanks{E-mail:
    hugorc@bison.ph.bham.ac.uk}, W.~J.~Chaplin$^{1,2}$, S.~Basu$^{3}$,
  A.~Serenelli$^{4}$, A.~Miglio$^{1,2}$ and D.~R.~Reese$^{1,2}$\\ $^1$
  School of Physics \& Astronomy, University of Birmingham, Edgbaston,
  Birmingham, B15 2TT, UK\\ $^2$ Stellar Astrophysics Centre (SAC),
  Department of Physics and Astronomy, Aarhus University, Ny Munkegade
  120, DK-8000 Aarhus C, Denmark\\ $^3$ Department of Physics and
  Astronomy, Yale University, P.O. Box 208101, New Haven, CT, 06520,
  USA\\ $^4$ Instituto de Ciencias del Espacio (ICE-CSIC/IEEC) Campus
  UAB, Carrer de Can Magrans, s/n 08193 Cerdanyola del Vall\'es, Spain}

\begin{document}

\maketitle

\begin{abstract}

Large-scale analyses of stellar samples comprised of cool, solar-like
oscillators now commonly utilize the so-called asteroseismic scaling
relations to estimate fundamental stellar properties. In this paper we
present a test of the scaling relation for the global asteroseismic
parameter $\nu_{\rm max}$, the frequency at which a solar-like
oscillator presents its strongest observed pulsation amplitude. The
classic relation assumes that this characteristic frequency scales
with a particular combination of surface gravity and effective
temperature that also describes the dependence of the cut-off
frequency for acoustic waves in an isothermal atmosphere, i.e.,
$\nu_{\rm max} \propto gT_{\rm eff}^{-1/2}$. We test how well the
oscillations of cool main-sequence and sub-giant stars adhere to this
relation, using a sample of asteroseismic targets observed by the NASA
\emph{Kepler} Mission. Our results, which come from a grid-based
analysis, rule out departures from the classic $gT_{\rm eff}^{-1/2}$
scaling dependence at the level of $\simeq 1.5\,\rm per cent$ over the
full $\simeq 1560\,\rm K$ range in $T_{\rm eff}$ that we tested. There
is some uncertainty over the absolute calibration of the scaling.
However, any variation with $T_{\rm eff}$ is evidently small, with
limits similar to those above.

\end{abstract}


 \section{Introduction}
 \label{sec:intro}

Cool main-sequence and sub-giant (i.e., solar-type) stars display
solar-like oscillations, pulsations that are stochastically excited
and intrinsically damped by near-surface convection. The solar-like
excitation mechanism gives rise to a rich spectrum of detectable
oscillations.  Data on global (or average) asteroseismic parameters
associated with the observed, rich oscillations spectra provide
important diagnostics of fundamental stellar properties. Extracting
these global parameters from oscillation spectra is usually very
straightforward, meaning a large number of stars can easily be
analysed. This can be particularly advantagous when low S/N ratios
either obstruct or prevent a more detailed analysis of individual
modes in the spectrum. Moreover, detailed analysis of stars with high
S/N or complicated oscillation spectra can be very time consuming.

Two global asteroseismic parameters have received particular
attention, and wide use.  One parameter is the average large frequency
separation, $\Delta\nu$. It is an average of the observed frequency
spacings between consecutive overtones $n$ of the same angular
(spherical) degree, $l$. The average large separation scales to a very
good approximation as $\rho^{1/2}$, where $\rho \propto M/R^3$ is the
mean density of a star of mass $M$ and surface radius $R$ (e.g. see
Tassoul 1980, Ulrich 1986, Christensen-Dalsgaard 1993). This scaling
has physical justification (e.g., see Reese et al. 2012).  The second
parameter, $\nu_{\rm max}$, is the frequency at which detected modes
of a star show the strongest amplitude. Brown et al. (1991) speculated
that this characteristic frequency might scale with the atmospheric
cut-off frequency for acoustic waves, $\nu_{\rm ac}$. Adopting an
isothermal approximation to the full equation describing the cut-off
frequency, one derives the scaling $\nu_{\rm ac} \propto \nu_{\rm max}
\propto gT_{\rm eff}^{-1/2}$ (Brown et al. 1991, Kjeldsen \& Bedding
1995). Numerous studies have shown empirically that $\nu_{\rm max}$
appears to follow this relation to reasonable approximation (e.g.,
Chaplin \& Miglio 2013 and references therein). While some headway has
been made on understanding the exact form of the scaling (e.g.,
Belkacem et al. 2011), a full theoretical justification remains
elusive. Additions to the scaling, for example a dependence on the
changing Mach number of the near-surface turbulent flow (Belkacem et
al. 2013), have also been proposed.

Information encoded in $\Delta\nu$ and $\nu_{\rm max}$ is commonly
exploited by using the above scaling dependencies normalized to
observed solar parameters or properties, i.e.,
  \begin{equation} 
  \Delta\nu \simeq \left(\frac{M}{\rm M_{\odot}} \right)^{1/2}
  \left(\frac{R}{\rm R_{\odot}} \right)^{-3/2}\, \Delta \nu _{\odot}
  \label{eq:dnu}
  \end{equation}
and
  \begin{equation} 
  \nu_{\rm max} \simeq \left(\frac{M}{\rm M_{\odot}}\right)
  \left(\frac{R}{\rm R_{\odot}}\right)^{-2} \left(\frac{T_{\rm
      eff}}{\rm T_{\rm eff\,\odot}}\right)^{-1/2}\, \nu_{\rm
    max\,\odot}.
  \label{eq:numax}
  \end{equation}
Typical solar values for the seismic parameters (e.g., see Chaplin et
al. 2014) are $\Delta\nu_{\odot} = 135.1\,\rm \mu Hz$ and $\nu_{\rm
  max\,\odot} = 3090\,\rm \mu Hz$. These solar-calibrated scaling
relations represent two equations in two unknowns when an estimate of
$T_{\rm eff}$ is also available, allowing us to solve directly for $M$
and $R$ to give so-called ``direct'' estimates of the stellar
properties. Moreover, each relation can be used independently to
provide direct estimates of the mean stellar density $\rho$
(Equation~\ref{eq:dnu}) or the surface gravity $g$
(Equation~\ref{eq:numax}). Alternatively, the relations may be
utilized (together or separately) as part of a grid-based search
code. Here, one searches amongst a grid of stellar evolutionary models
to find those models whose predicted asteroseismic or atmospheric
parameters match the actual observed parameters (at the level of the
observational uncertainties). The solar-calibrated relations provide
the means to translate model properties ($R$, $M$, $T_{\rm eff}$) to
expected values for $\Delta\nu$ and $\nu_{\rm max}$, thereby allowing
a comparison to be made with the observations.

These global asteroseismic parameters and their associated scaling
relations are now being employed in analyses of large samples of
solar-like oscillators to, for example, generate catalogues of
asteroseismic stellar properties (e.g., see Huber et al. 2013, Chaplin
et al. 2014, Casagrande et al. 2014, Pinsonneault et al. 2014).  Tests
in the literature of the scaling relations for solar-type stars have,
by and large, returned encouraging results.  Studies have most
commonly looked at data on estimated stellar radii, and include
comparisons with very accurate properties from binaries, parallaxes
and long baseline interferometry (e.g., Bruntt et al. 2010, Bedding
2011, Miglio 2012, Huber et al. 2012, Silva Aguirre et al. 2012, White
et al. 2013). Results have tested the combination of the two scaling
relations to levels of around 4\,per cent in inferred radii, and
10\,per cent in inferred masses (Chaplin et al. 2014).

Results on red giants are more complicated. For example, He-core
burning and H-shell burning giants with the same mass and radius can
have a different $\Delta\nu$, due to differences in the sound-speed
profile in the outer layers (Miglio et al. 2012), which implies a
different absolute scaling for Equation~\ref{eq:dnu}. Meanwhile, other
studies have looked at open clusters in the \emph{Kepler} field,
comparing results on red giants inferred from asteroseismology and
from turnoff eclipsing binaries (Brogaard et al. 2012, Sandquist et
al. 2013).

Our goal in this paper it to test, empirically, the accuracy of the
classic $\nu_{\rm max}$ scaling relation for oscillations seen in
solar-type stars, i.e., cool main-sequence and sub-giant stars (we
leave a study of giants to future work).  There is actually very
little in the literature on the $\nu_{\rm max}$ scaling alone, which
partly reflects the difficulty of obtaining the data needed to test
the one relation in isolation. A recent example made use of
interferometric data on a few very bright \emph{Kepler} targets: White
et al. (2013) concluded that results on the F-type star $\theta$\,Cyg
may point to problems for the $\nu_{\rm max}$ scaling in the hottest
solar-type stars.

Our basic approach is as follows. We use data on a sample of around
500 stars observed by the NASA \emph{Kepler} Mission (the same sample
as in Chaplin et al. 2014).  Each star in our sample has a measured
$\nu_{\rm max}$, which comes from analysis of the \emph{Kepler}
data. Throughout the rest of the paper we shall refer to these actual,
measured values as $\nu_{\rm max}({\rm data})$. Now, the classic
scaling relation (Equation~\ref{eq:numax}) gives $\nu_{\rm max}$ in
terms of $g$ and $T_{\rm eff}$, or, to be more specific, the
combination $g\,T_{\rm eff}^{-1/2}$. We can therefore test the scaling
if we have independent measures of
  \begin{equation} 
  \nu_{\rm max}({\rm grid}) \equiv \left(\frac{g}{\rm g_{\odot}}\right)
  \left(\frac{T_{\rm eff}}{\rm T_{\rm eff\,\odot}}\right)^{-1/2}\,
  \nu_{\rm max\,\odot},
  \label{eq:nug}
  \end{equation}
to which the observed $\nu_{\rm max}({\rm data})$ may be compared. The
$\nu_{\rm max}({\rm grid})$ are so-named because we adopt a grid-based
search technique to estimate them, using as inputs the asteroseismic
average large separations, $\Delta\nu$, of the stars along with
photometric temperatures, $T_{\rm eff}$, derived using the Infrared
Flux Method (IRFM) and, where available, metallicities [Fe/H] from
spectroscopy.  We search grids of stellar evolutionary models to find
those models whose predicted $\{ \Delta\nu, T_{\rm eff}, {\rm [Fe/H]}
\}$ match the actual observed inputs. Each model in the grid also has
a computed $\nu_{\rm max}({\rm grid})$, which comes from its $M$, $R$
and $T_{\rm eff}$. The best-matching models will have the most likely
values of $\nu_{\rm max}({\rm grid})$. A suitable, likelihood-weighted
average therefore provides an estimate of $\nu_{\rm max}({\rm grid})$
for every star (see Section~\ref{sec:pipes}), which may then be
compared directly to the observed $\nu_{\rm max}({\rm data})$. Any
departures from a one-to-one correspondance of the values would point
to problems with the classic scaling relation, and allow us to
quantify departures from the scaling. In sum, we leverage the
potential of using asteroseismic results on a large number of
solar-type stars to follow a statistical (ensemble) approach to the
analysis and to thereby beat-down the errors. The approach is not
dissimilar to that adopted by Morel et al. (2014) for part of their
analysis of red giants observed by the CNES/ESA CoRoT Mission, which
compared values of the surface gravity estimated using $\Delta\nu$ on
the one hand and $\nu_{\rm max}$ on the other.

The layout of the rest of the paper is as follows. We begin in
Section~\ref{sec:method} with a discussion of the basic methodology
adopted to test the $\nu_{\rm max}$ scaling
relation. Section~\ref{sec:dataandgrid} then introduces the real
\emph{Kepler} data, the artificial data we made to check our
methodology (Section~\ref{sec:data}), and the grid-based search
pipelines used to estimate the $\nu_{\rm max}({\rm grid})$ values
(Section~\ref{sec:pipes}). Our results are presented in
Section~\ref{sec:res}, beginning with a validation of the methodology
using artificial data (Section~\ref{sec:artres}), followed by results
from the actual \emph{Kepler} sample. We finish in
Section~\ref{sec:conc} with concluding remarks.

 \section{Method}
 \label{sec:method}

We may in principle use some suitable grid-based results to test the
$\nu_{\rm max}$ scaling relation, albeit with caveats that we will
discuss and address below.  Let us suppose for the moment that
grid-based searches using the set of inputs
 \[
 \{ \Delta\nu, T_{\rm eff}, {\rm [Fe/H]} \}
 \]
provide robust, unbiased estimates of the combination $gT_{\rm
  eff}^{-1/2}$, as calibrated to give $\nu_{\rm max}({\rm grid})$
defined by Equation~\ref{eq:nug} above.  Assuming the temperatures and
metallicities to be unbiased, at least to a level that will not
influence significantly estimation of the combination $gT_{\rm
  eff}^{-1/2}$, the fractional differences
\[
\left[ \nu_{\rm max}({\rm data})/\nu_{\rm max}({\rm grid})\right] -1,
\]
will provide a direct estimate of the bias in the $\nu_{\rm max}$
scaling, i.e., the fractional amount by which the $\nu_{\rm max}({\rm
  data})$ values are over or underestimated relative to $\left(g/{\rm
  g_{\odot}}\right)\left(T_{\rm eff}/{\rm T_{\rm
    eff\,\odot}}\right)^{-1/2}\nu_{\rm max\,\odot}$.

Crucial to the approach is the accuracy of $\nu_{\rm max}({\rm
  grid})$.  First, we know to expect a small bias if the grid-based
search pipelines employ the $\Delta\nu$ scaling relation, as we now go
on to explain.  When the scaling relation is used in the grid-based
searches, the fundamental properties of models in the grid are used as
inputs to Equation~\ref{eq:dnu} to yield model estimates of
$\Delta\nu$ for comparison with the observed
separations. Alternatively, one may circumvent use of the scaling by
computing for each model a set of theoretical oscillation frequencies
(e.g., radial-mode frequencies spanning the same orders as those
observed in the real data), from which one may then estimate the
required $\Delta\nu$ from a suitable fit to those frequencies.

It is now well known that predictions made by the calibrated
scaling-relation (Equation~\ref{eq:dnu}) have small, systematic
differences with respect to predictions from model-computed
frequencies (e.g., see Ulrich 1986, White et al. 2011; Mosser et
al. 2013). For solar-type stars, these differences can be up to
$\simeq 2\,\rm per cent$ in size, and become more pronounced at
effective temperatures progressively further away from $T_{\rm eff}
\simeq 5700\,\rm K$ (e.g., see Figures 5 and 6 of White et al. 2011;
and figures in Chaplin et al. 2014). Here, we have employed grid-based
search pipelines that can run with or without the $\Delta\nu$ scaling
relation. We use one pipeline that may be run either using individual
model-calculated frequencies or the $\Delta\nu$ scaling relation; and,
for comparison, two other pipelines that used the $\Delta\nu$ scaling
only.

Second, we must consider the impact of the poor modelling of the
near-surface layers of stars. In the case of the Sun it is now well
established that this gives rise to a frequency-dependent offset
between observed and model-calculated oscillation frequencies (e.g.,
see Chaplin \& Miglio 2013, and references therein). This so-called
``surface term'' increases in magnitude with increasing overtone
number, $n$.  The amount by which $\Delta\nu$ is affected will depend
on the variation of the surface term with $n$. Tests of the grid-based
method (Basu et al. 2010) indicate that the impact of the solar
surface term on $\Delta\nu$ -- which decreases the observed solar
$\Delta\nu$ by just under 1\,per cent compared to model predictions --
leads only to very small errors in the inferred solar properties,
certainly well within the observational uncertainties associated with
the \emph{Kepler} data used in this paper. Stellar surface terms would
need to be substantially larger than the solar term to produce
significant bias in our results. However, the nature of the term in
other stars remains rather poorly understood, and so this caveat
should be borne in mind.

Our tests with artificial data do nevertheless provide some insights
on the sensitivity of the results to such offsets. The artificial data
come from models computed using a different stellar evolutionary code
and different input physics to those of the grids to which the
grid-based pipelines are coupled. This can give rise to
``surface-term'' like offsets between models in the artificial sample
in the grids that share the same fundamental properties (in particular
from differences in boundary conditions, matching to model atmospheres
etc.).

 \section{Data and grid pipelines}
 \label{sec:dataandgrid}

 \subsection{Real and artificial data}
 \label{sec:data}

The observational data for our study come from Chaplin et al. (2014).
This study produced an asteroseismic catalogue from an extensive
grid-based analysis of more than 500 solar-like oscillators, which
were observed by \emph{Kepler} as part of an asteroseismic survey that
was conducted over the first 10\,months of science operations.
Stellar properties were estimated from a grid-based analysis using the
global asteroseismic parameters $\Delta\nu$ and $\nu_{\rm max}$
together with complementary photometric and spectroscopic data as the
inputs. Homogeneous sets of effective temperatures $T_{\rm eff}$ were
available for the full sample of stars, courtesy of complementary
ground-based photometry.  A homogeneous set of spectroscopic
parameters ($T_{\rm eff}$ and [Fe/H]) was also available from Bruntt
et al. (2012), but only for a subset of 87 stars in the sample

Here, we make use of the global asteroseismic parameters and the
complementary data to perform the new grid-based analysis needed to
test the $\nu_{\rm max}$ scaling relation.  The Chaplin et al. (2014)
sample is dominated by cool main-sequence and sub-giant stars but does
contain a small fraction of stars at the base of the red-giant branch
that were serendipitously observed as part of the short-cadence
asteroseismic survey. We have removed the more evolved stars from the
sample (which will be the subject of a separate study). The selected
sample contains 426 solar-type stars.

As noted above, complementary photometry was available on the entire
sample. This allowed us to perform a new, homogenous grid-based
analysis on all the selected solar-type stars, but at the cost of not
having robust, well-constrained estimates of [Fe/H] for each star
since the complementary photometry available to us in the
\emph{Kepler} Input Catalogue (KIC; see Brown et al. 2011) does not
provide strong constraints on metallicity.  Just like Chaplin et
al. (2014), we therefore adopted an average [Fe/H] value as input for
every star when we analysed the 426-star sample. We actually tried two
different values: One set of results came from using the average of
the 87 metallicities measured by Bruntt et al. (2012), i.e., ${\rm
  [Fe/H]}=-0.05\,\rm dex$; while the other results came from using
${\rm [Fe/H]}=-0.20\,\rm dex$ for all stars (e.g., see Silva Aguirre
et al. 2011), the value adopted in Chaplin et al. (2014). In both
cases we adopted large input uncertainties of $\pm 0.3\,\rm dex$.

In spite of the weak constraints on [Fe/H], we still obtained more
precise results from the larger 426-star sample having complementary
photometric data than we did from the smaller sample with
complementary spectroscopic data because the larger sample size
compensated for the inferior precision in [Fe/H]. Results obtained
were similar, and hence in what follows we present detailed results
from the photometric sample. This sample also provided much better
coverage in the domain where $T_{\rm eff} > 6000\,\rm K$.

Finally with regards to the input data, we note that the photometric
temperatures were the Infrared Flux Method (IRFM) estimates from
Chaplin et al. (2014), which were calculated using multi-band
photometry in the Two Micron All Sky Survey (2MASS; see Skrutskie et
al. 2006) JHK bands, and in the Sloan Digital Sky Survey (SDSS) $griz$
bands (both available in the KIC).


\begin{figure*}

 \centerline {\epsfxsize=9.0cm\epsfbox{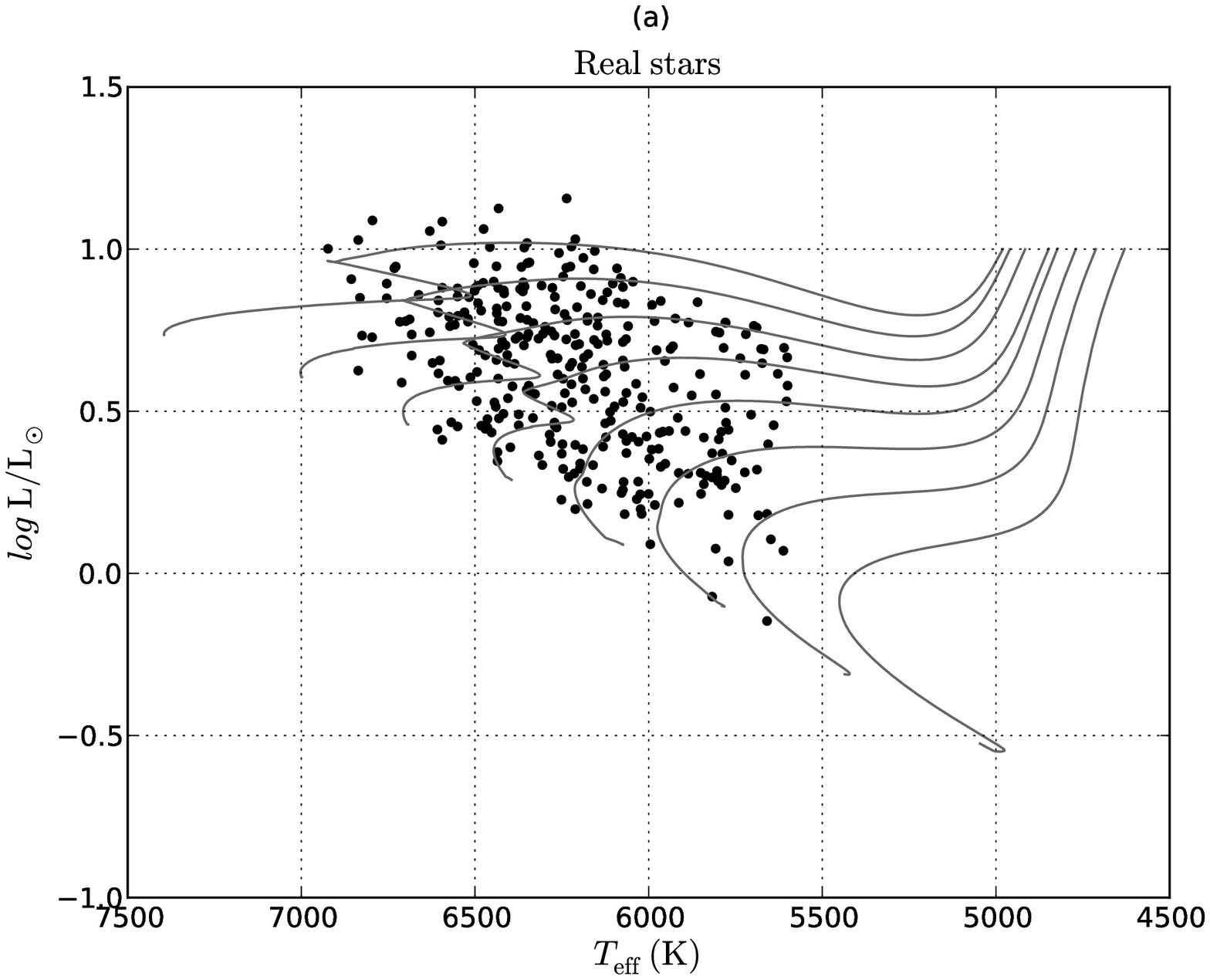}
              \epsfxsize=9.0cm\epsfbox{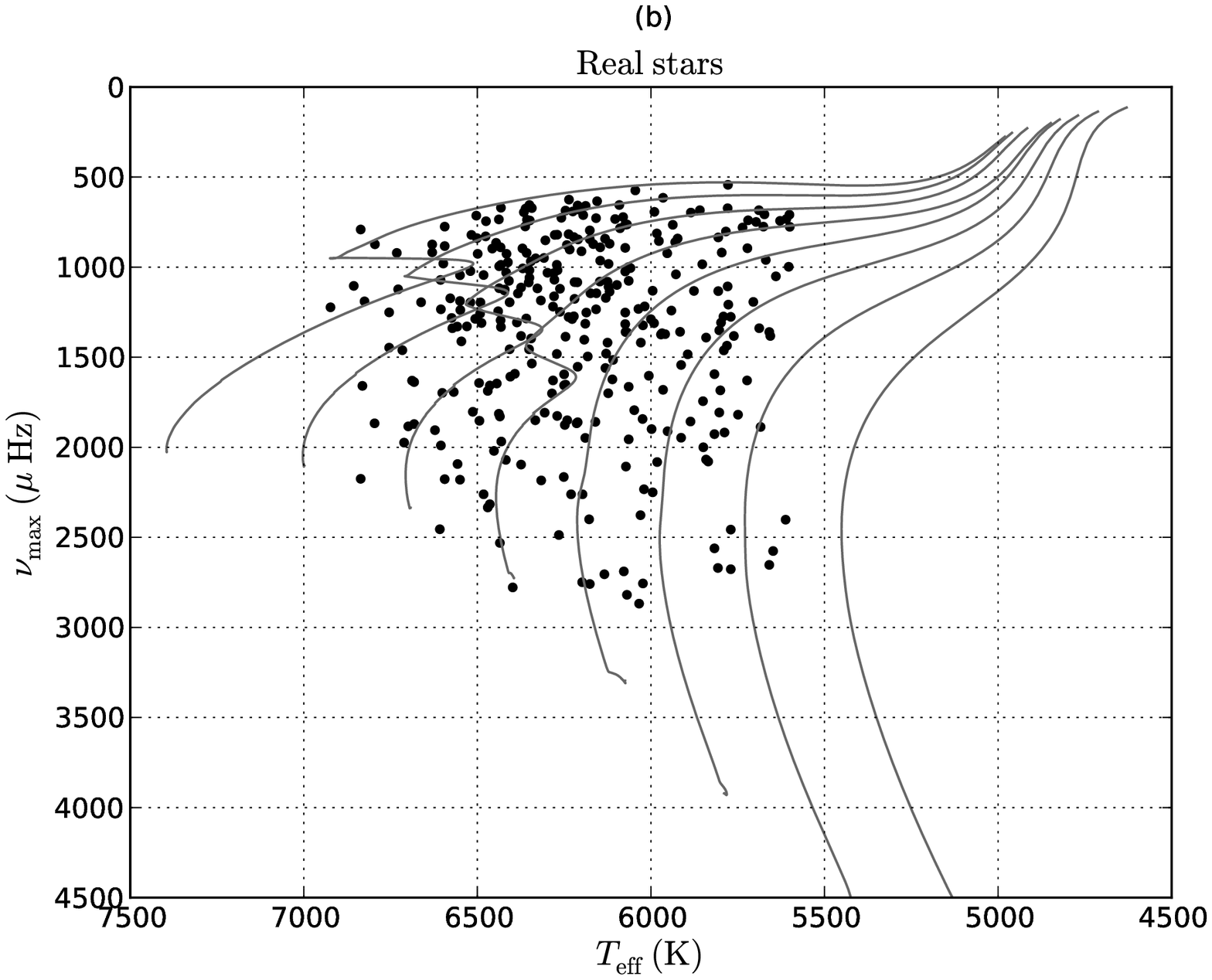}}
 \centerline {\epsfxsize=9.0cm\epsfbox{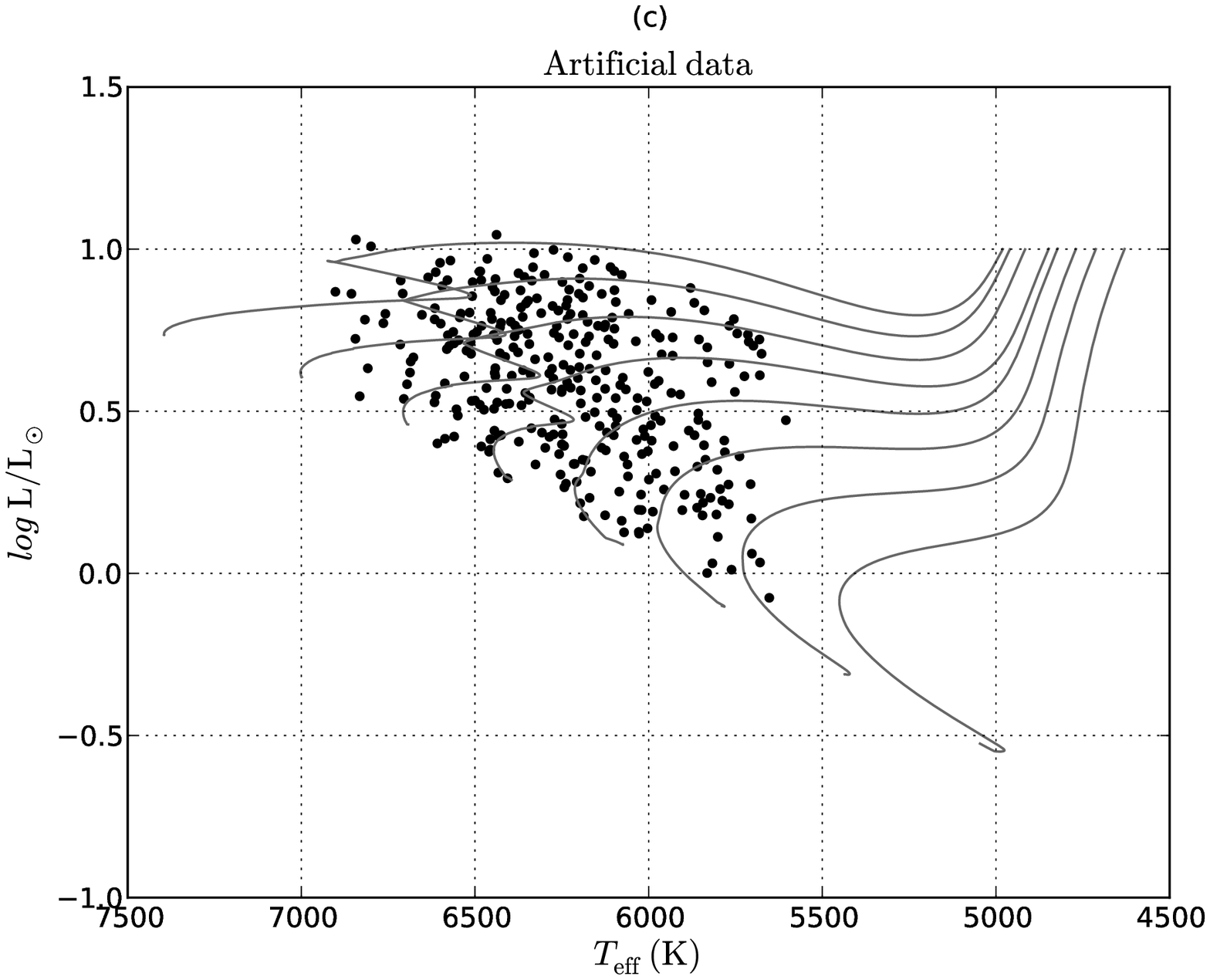}}

 \caption{Panel (a): Hertzsprung-Russell diagram of the asteroseismic
   sample of \emph{Kepler} solar-type stars.  Panel (b): measured
   values, $\nu_{\rm max}({\rm data})$, from analysis of the
   oscillation spectra of the real stars, also as a function of
   effective temperature.  Panel (c): Hertzsprung-Russell diagram of
   the sample of artificial targets, computed from MESA models. The
   lines in the panels show evolutionary tracks computed by the MESA
   code (see text), for models having solar composition and masses
   ranging from 0.8 to $1.5\,\rm M_{\odot}$ (in steps of $0.1\,\rm
   M_{\odot}$).}

 \label{fig:HR}
\end{figure*}


Panel (a) of Fig.~\ref{fig:HR} plots the locations of the sample of
selected \emph{Kepler} solar-type stars on a Hertzsprung-Russell
diagram.  The lines mark evolutionary tracks computed by the Modules
for Experiments in Stellar Astrophysics (MESA) code (Paxton et
al. 2011; see below), for models having solar composition and masses
ranging from 0.8 to $1.5\,\rm M_{\odot}$ (in steps of $0.1\,\rm
M_{\odot}$).  Panel (b) of the same figure shows the measured
$\nu_{\rm max}({\rm data})$ of the \emph{Kepler} sample, from analysis
of the oscillation spectra of the stars, indicating the parameter
range tested by the analysis.

We also used artificial data to test and validate our methodology.
The artificial sample of stars was comprised of models drawn from an
evolutionary grid, computed using MESA. The grid spanned the range
$0.8 \le M/{\rm M_{\odot}} \le 1.5$ and $-0.8 \le {\rm [Fe/H]} \le
0.8$ (both in steps of 0.1), with tracks computed from the pre-main
sequence to the base of the red-giant branch.

The Grevesse \& Noels (1993) value of $Z_{\odot}/X_{\odot} = 0.0245$
was used to translate between model values of [Fe/H] and $Z/X$.  The
OPAL equation of state (Rogers \& Nayfonov 2002) and OPAL opacities
(Iglesias \& Rogers 1996) were used, augmented by low-temperature
opacities from Ferguson et al. (2005). Nuclear reaction rates were
from NACRE (Angulo et al. 1999), with updates for the $^{14}{\rm
  N}(p,\gamma)^{15}{\rm O}$ (Imbriani et al. 2004, 2005)
reaction. Convection was treated according to mixing-length theory,
using the solar-calibrated mixing length parameter.  Diffusion and
effects of rotational mixing were not included.  The primordial Helium
abundance was fixed to $Y_{\rm p}=0.2484$ (Cyburt et al.  2003), and
the helium enrichment set to $\Delta Y / \Delta Z = 2$ (e.g., Chiosi
\& Matteucci 1982, Carigi \& Peimbert 2008). We refer the reader to
Paxton et al. (2011) for further details.

Artificial stars were drawn from the grid by seeking a ``best
matching'' model for each of the solar-type stars in the \emph{Kepler}
sample (via a $\chi^2$ minimization).  We found that a selection based
solely on a comparison of the observed (\emph{Kepler}) and model
(grid) values of $T_{\rm eff}$ and $\Delta\nu$ (the latter using the
calibrated scaling relation) was sufficient to produce an artificial
sample that had distributions in each of the fundamental properties
that were a reasonable match to those for the real sample. Panel (c)
of Fig.~\ref{fig:HR} marks the locations of the selected models in our
artificial sample of stars (using the pristine model
parameters). Owing to the limited resolution of the grid, we found that
some real stars had the same best-matching artificial model. This
meant that our final sample of selected artificial stars was comprised
of 306 unique models.

Next, we computed adiabatic oscillation frequencies for each of the
306 selected models, using the GYRE (Townsend \& Teitler 2013) stellar
oscillations code. The $\Delta\nu$ of each artificial star was then
given by the best-fitting gradient of a linear fit to the radial
order, $n$, of the five $l=0$ frequencies centred on the estimated
$\nu_{\rm max}$ (see below) of the model. This approach gives values
that are representative of the average values extracted from the real
data.  We also tested the impact of calculating $\Delta\nu$ using a
different number of orders (to reflect the varying data quality and
S/N levels in the real sample of stars) and from adopting a weighted
fit of the frequencies (to reflect the impact of the changing S/N in
any given observed spectrum). Neither change had a significant impact
on our results.

We computed three different sets of $\nu_{\rm max}$ for the artificial
sample. One set was computed assuming perfect adherence to the
solar-calibrated scaling relation, i.e., by using
Equation~\ref{eq:numax}. We made two other sets by applying a
temperature-dependent fractional offset to the solar-calibrated
scaling relation, i.e., a computation that took the form
  \begin{equation} 
  \nu_{\rm max} = {\cal F}(T_{\rm eff})
  \left(\frac{M}{\rm M_{\odot}}\right) \left(\frac{R}{\rm
    R_{\odot}}\right)^{-2} \left(\frac{T_{\rm eff}}{\rm T_{\rm
      eff\,\odot}}\right)^{-1/2}\, \nu_{\rm max\,\odot},
  \label{eq:off1}
  \end{equation}
with ${\cal F}(T_{\rm eff})$ being the fractional
temperature-dependent offset. We applied a linear and quadratic
offset, respectively, both of which are shown in panel (a) of
Fig.~\ref{fig:art3} (in the figure that also shows results from the
data; see below for further discussion).  Use of the first set of
``perfect'' values allowed us to test the impact of biases not
associated with the $\nu_{\rm max}$ scaling relation; while the other
sets allowed us to test whether we could recover information on a
known bias in the $\nu_{\rm max}$ scaling.

Finally, with each artificial star then having calculated values of
$\Delta\nu$ and $\nu_{\rm max}$, and a model $T_{\rm eff}$, it
remained to add noise and to assign uncertainties to those parameters
for input to the grid pipelines. This meant that in our analysis we
would treat the artificial data in exactly the same way as the real
data. Here, we simply used the relevant parameter uncertainties of the
real \emph{Kepler} star to which each artificial star was
associated. To make a given realization of the artificial datasets, we
added Gaussian noise to each pristine input parameter, multiplied by
the relevant parameter uncertainty, to give $\nu_{\rm max}({\rm
  data})$, $\Delta\nu({\rm data})$, and $T_{\rm eff}({\rm data})$ for
each artificial star.

 \subsection{Grid pipelines}
 \label{sec:pipes}

We utilized three different grid-based pipeline codes to return
estimates of the parameter $\nu_{\rm max}({\rm grid})$ for all stars
in the real and artificial samples:
 \begin{itemize}
 \item[--] the Bellaterra Stellar Properties Pipeline (BeSPP)
   (Serenelli et al. 2013, extended for asteroseismic modelling);
 \item[--] the Yale-Birmingham (YB) (Basu et al. 2010, 2012, Gai et
   al. 2011); and
 \item[--] PARAM (da Silva et al. 2006; Miglio et al. 2013; Rodrigues
   et al. 2014);
 \end{itemize}
The BeSPP pipeline was run with a grid comprised of models constructed
with the GARSTEC code (Weiss \& Schlattl 2008). The parameters of the
grid are described in Silva Aguirre et al. (2012). BeSPP was run in
two different modes of operation: one where grid-model estimates of
$\Delta\nu$ were computed using adiabatic oscillation frequencies
(frequency mode); and one where the estimates were instead computed
using the solar-calibrated scaling relation (Equation~\ref{eq:dnu};
scaling-relation mode). The other two pipelines were run only in the
latter, scaling-relation mode.  PARAM was run using a grid comprising
models made by the Padova group (Marigo et al. 2008). Further details
may be found in Miglio et al. (2013).  The YB pipeline returned
results using five different sets of stellar models: grids computed by
the Dartmouth (Dotter et al. 2008) and Padova groups (Bressan et
al. 2012); the set of YY isochrones (Demarque et al.  2004); a grid
comprised of the BASTI models of Pietrinferni et al. (2004), computed
for use in asteroseismic studies (see Silva Aguirre et al 2013); and,
finally, a grid constructed using the YREC Code (Demarque et
al. 2008), which is described by Basu et al. (2012). This grid has
been used in other papers, and we retain the YREC2 name here.

We also report results from YB which are labelled ALL. This set of
results was generated by combining the YB analysis over all five grids
to compute what are in essence averages from a composite ``super
distribution''. This is a new addition to the YB code that has not
previously been documented in the literature, and so we provide
further details here.

The YB pipeline determines the properties of a star using the given
observational input (central) parameter set. A key step in the method
is to generate 10,000 input parameter sets by adding different random
realisations of Gaussian noise (commensurate with the input
uncertainties) to the actual (central) observational input parameter
set. For each realisation, we find all models in a grid within
3$\sigma$ of the input uncertainties, and use these models to define a
likelihood function (e.g., see Gai et al. 2011, Basu et al. 2012 for
details).  The estimated property, e.g., $\nu_{\rm max}({\rm grid})$,
is then the likelihood-weighted average of the property of the
selected models. The 10001 values of any given property estimated from
the central value and the 10000 realisations, form the probability
distribution function for that parameter.  In the YB pipeline we adopt
the median of the distribution as the estimated value of the property,
and we use 1$\sigma$ limits from the median as a measure of the
uncertainties. The ALL results were obtained by constructing a
consolidated probability distribution for a given star by adding
together the five different distribution functions obtained using the
five different grids. Then we determined the median of this
consolidated or ``super'' distribution function, to determine the
average $\nu_{\rm max}({\rm grid})$.

All three pipelines were employed in the grid-based analysis described
in Chaplin et al. (2014), where summary details of the physics
employed in the grids may also be found.

 \section{Results}
 \label{sec:res}

 \subsection{Results from artificial data}
 \label{sec:artres}

\begin{figure*}

 \centerline {\epsfxsize=9.0cm\epsfbox{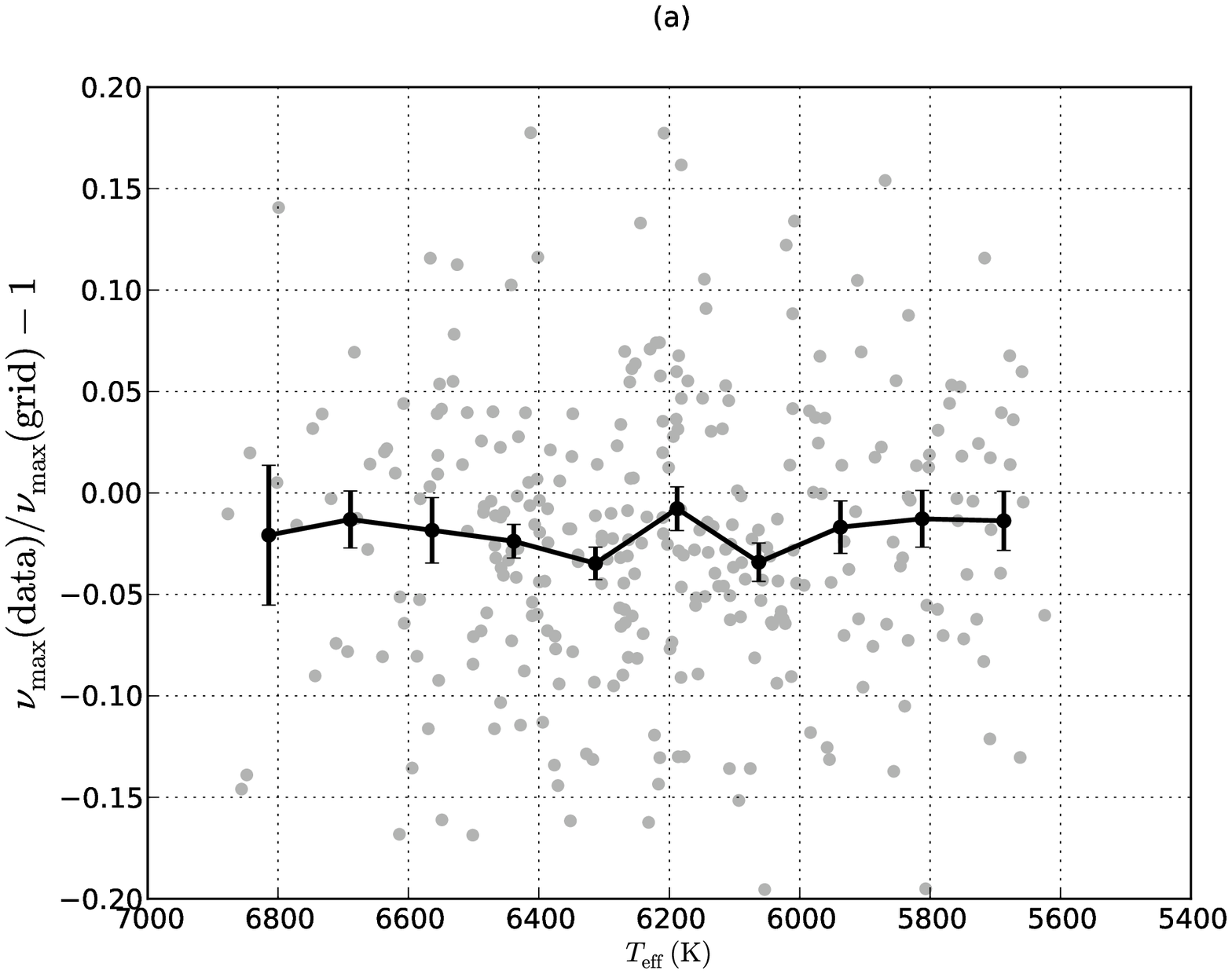}
              \epsfxsize=9.0cm\epsfbox{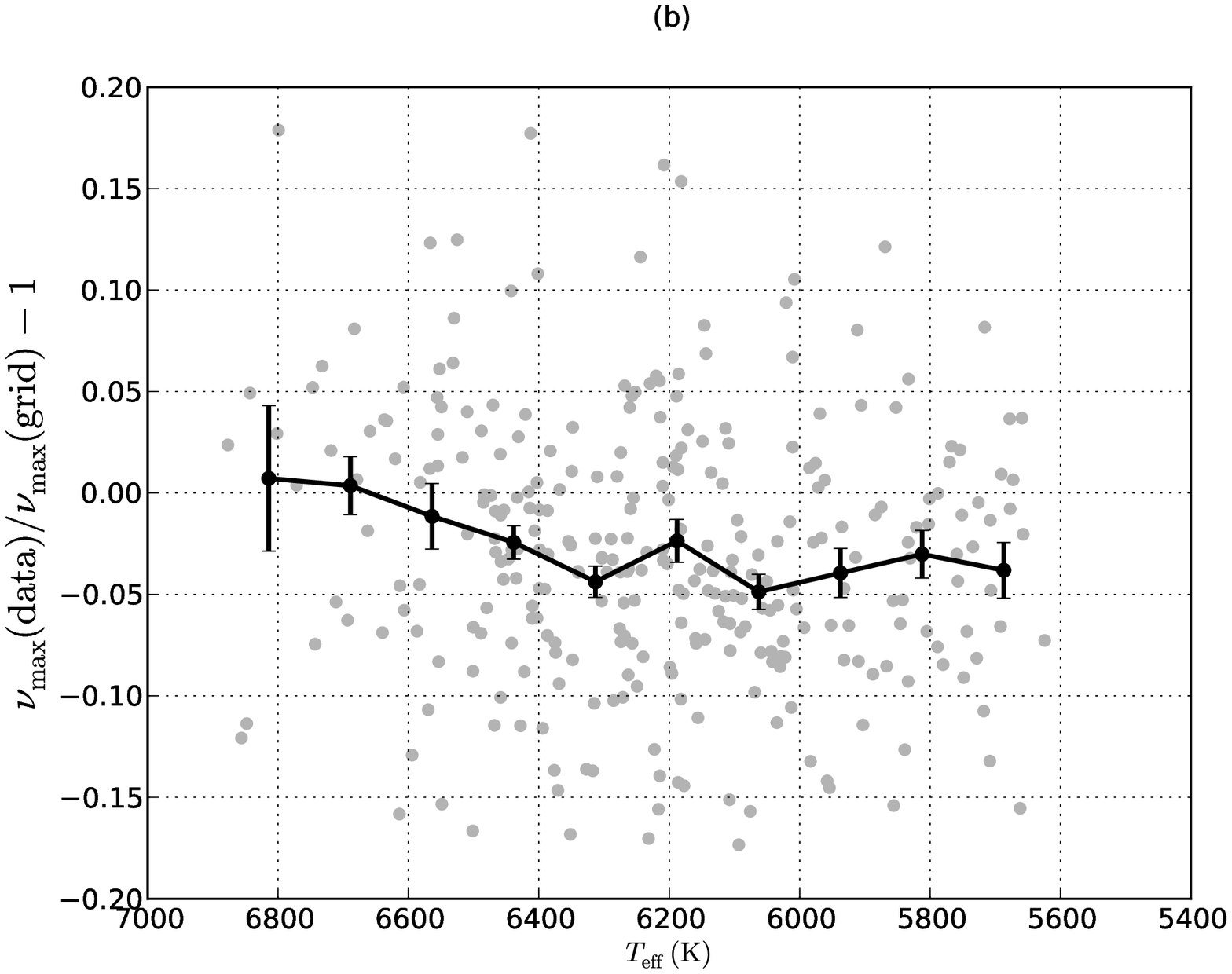}}
 \centerline {\epsfxsize=9.0cm\epsfbox{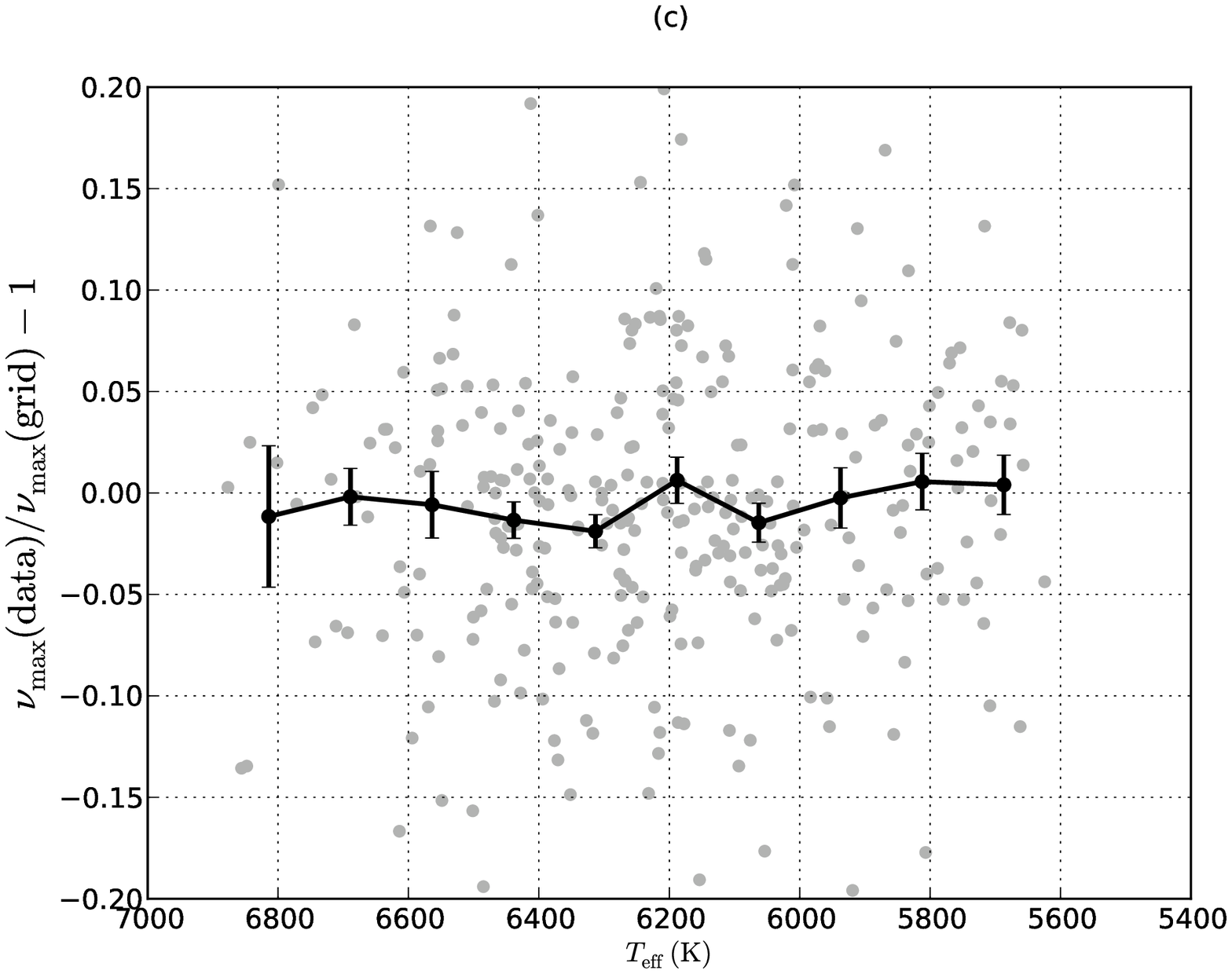}}

 \caption{Results from BeSPP grid pipeline, for artificial data
   following perfect adherence to the $\nu_{\rm max}$ scaling
   relation. Panel (a) shows results from when BeSPP is run in
   frequency mode; while panel (b) shows results when the pipeline is
   run in scaling-relation mode (see text).  Plotted are fractional
   differences $\nu_{\rm max}({\rm data})/\nu_{\rm max}({\rm
     grid})-1$, with grid-based estimates calculated assuming an input
   [Fe/H] of $-0.05\,\rm dex$ for every artificial star.  Results are
   plotted for individual artificial stars (symbols) and for averages
   computed over 130-K bins in $T_{\rm eff}$ (lines). The error bars
   mark uncertainties on each average. Panel (c): Frequency-mode
   results from analysing the same artificial data, but now with an
   input [Fe/H] of $-0.20\,\rm dex$.}

 \label{fig:art1}
\end{figure*}

We begin with results from analysing the first set of artificial
data. Recall that in this case $\nu_{\rm max}({\rm data})$ for each
artificial star -- essentially the proxy for what would be the
observed $\nu_{\rm max}$ of each real star -- was computed assuming
strict adherence to the solar-calibrated scaling relation.  The top
two panels of Fig.~\ref{fig:art1} show results from the BeSPP
pipeline, run in both frequency mode [panel (a)] and scaling-relation
mode [panel (b)], on a representative noise realization of the
artificial set. Both panels show fractional differences $\nu_{\rm
  max}({\rm data})/\nu_{\rm max}({\rm grid})-1$, i.e., fractional
differences between the simulated measurements, $\nu_{\rm max}({\rm
  data})$, and the grid-based estimates, $\nu_{\rm max}({\rm
  grid})$. Here, grid-based estimates were calculated assuming an
input [Fe/H] pegged to the Bruntt et al. average, i.e., ${\rm
  [Fe/H]}=-0.05\,\rm dex$.  Results are shown both for the individual
artificial stars (symbols) and for averages computed over 130-K bins
in $T_{\rm eff}$ (lines). The error bars mark uncertainties on each
average. The scatter in the fractional differences is entirely
consistent with the formal uncertainties on the differences (which are
propagated from the individual formal uncertainties on $\nu_{\rm
  max}({\rm data})$ and $\nu_{\rm max}({\rm grid})$).


The results on the whole show the expected trends. The
scaling-relation-mode results in panel (b) are not flat. The
differences show an upward trend with increasing $T{\rm eff}$, which
is due to the known offsets in the $\Delta\nu$ scaling. This tells us
that if there were no $\nu_{\rm max}$ bias in the real data we should
expect to see trends like these when using a scaling-relation-based
grid pipeline.

With regards to the frequency-mode results in the panel (a) of
Fig.~\ref{fig:art1}, as expected the upward trend from the $\Delta\nu$
scaling is absent, and we see a flat trend in the comparison at the
level of precision of the data. The frequency-mode results therefore
allow us to conclude, correctly, that the artificial data follow a
$gT_{\rm eff}^{-1/2}$ like scaling. That said, the differences do show
a small absolute offset, albeit one that does not change significantly
with $T_{\rm eff}$. The absolute offset has more than one
contribution.

First, there is a contribution due to the uncertainty in [Fe/H] we
adopted for the input data (to mimic that for the real 426-star
sample). Panel (c) of Fig.~\ref{fig:art1} shows results from BeSPP run
in frequency mode, but with ${\rm [Fe/H]}=-0.20\,\rm dex$ now used as
input for all stars. This change to the input [Fe/H] produces a shift
in the absolute offset of just under 1\,per cent. Second, there will
be a contribution to the absolute offset arising from differences
between the models on which the artificial stars are based and those
used in the BeSPP grid. These differences can give rise to what looks
like a ``surface term'' offset (see discussion in
Section~\ref{sec:method} above). We conclude that we should expect
there to be some uncertainty over the \emph{absolute} calibration of
the relation for results on the real \emph{Kepler} data. We shall come
back in Section~\ref{sec:realres} to attempt an estimate of the
relative contributions of the above effects to the uncertainty in the
absolute calibration.


\begin{figure*}

 \centerline {\epsfxsize=9.0cm\epsfbox{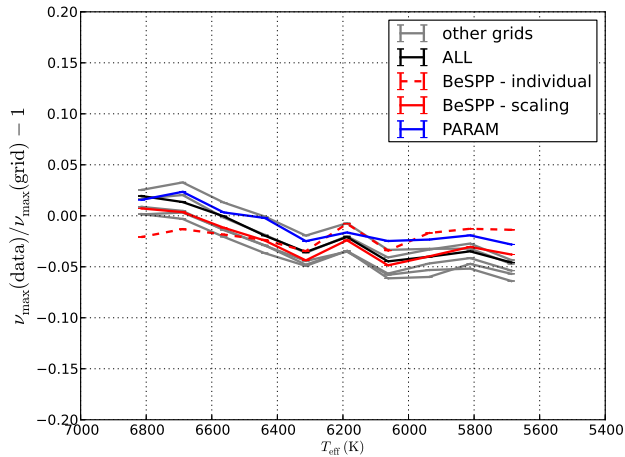}}

 \caption{Fractional differences $\nu_{\rm max}({\rm data}) /\nu_{\rm
     max}({\rm grid})-1$ returned by all three grid pipelines, for
   artificial data following perfect adherence to the solar-calibrated
   $\nu_{\rm max}$ scaling relation (Equation~\ref{eq:numax}).
   Different lines show 130-K averages in $T_{\rm eff}$ for different
   pipeline-grid combinations (see annotation), all of which used an
   input [Fe/H] of $-0.05\,\rm dex$.}

 \label{fig:art2}
\end{figure*}


Fig.~\ref{fig:art2} also includes results from the YB and PARAM
pipelines, which were coupled to a variety of grids with ${\rm
  [Fe/H]}=-0.05\,\rm dex$ used as input.  The YB ALL results are
plotted in black, the YB results from individual grids in grey, and
the PARAM results in blue.  Recall that these pipelines were run only
in scaling-relation mode. To aid the clarity of the plots we present
just the $T_{\rm eff}$-binned averages.  Results from the other
pipelines follow the BeSPP scaling-relation results from the top
panels of Fig.~\ref{fig:art1}, which we plot again here (in red) for
direct comparison.  The shapes of the scaling-relation trends for each
grid-pipeline combination are similar -- showing an upward trend with
increasing $T_{\rm eff}$, again due to the known offsets in the
$\Delta\nu$ scaling -- with an extreme spread of approximately
3.5\,per cent in the fractional differences.


\begin{figure*}

 \centerline {\epsfxsize=9.0cm\epsfbox{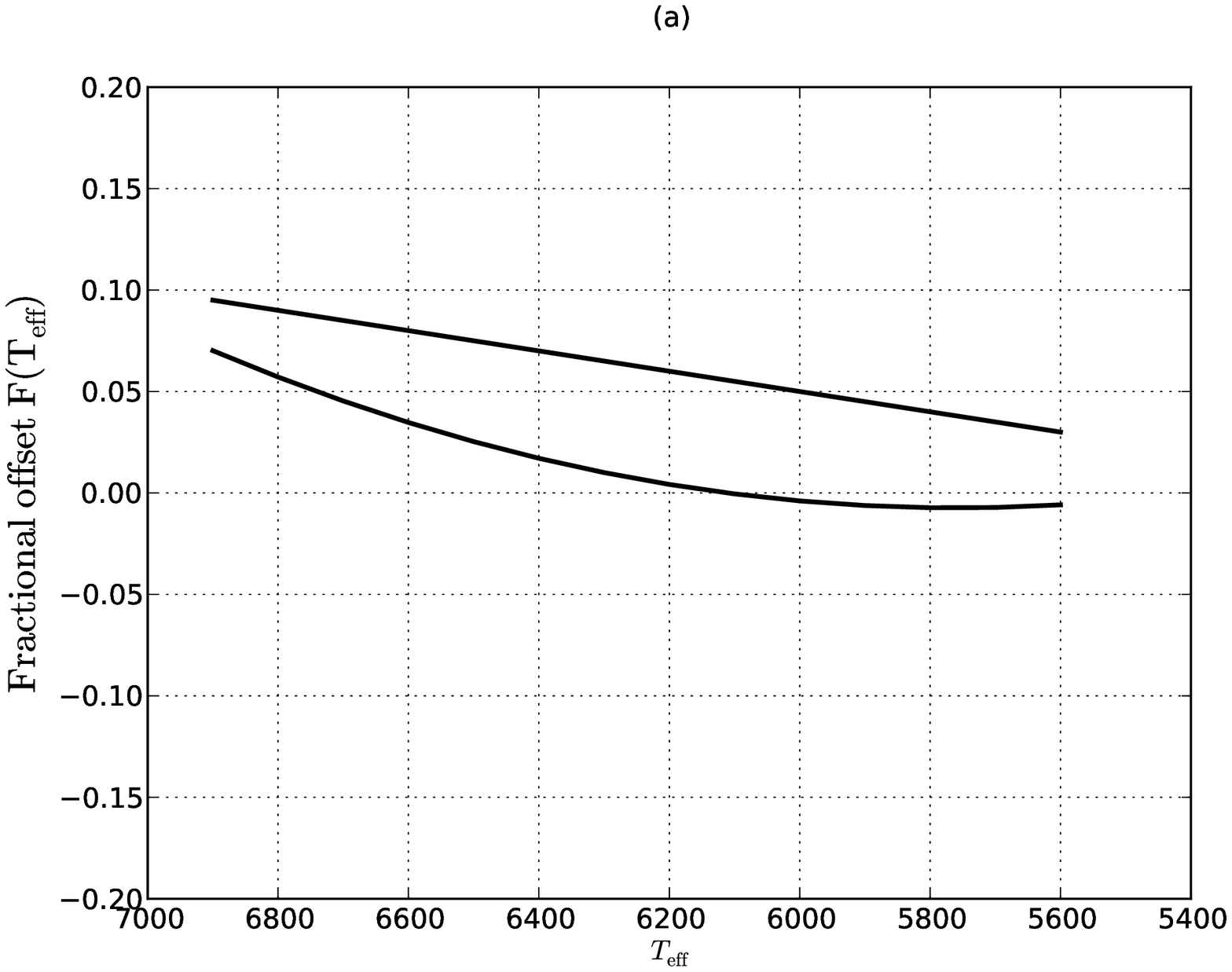}}
 \centerline {\epsfxsize=9.0cm\epsfbox{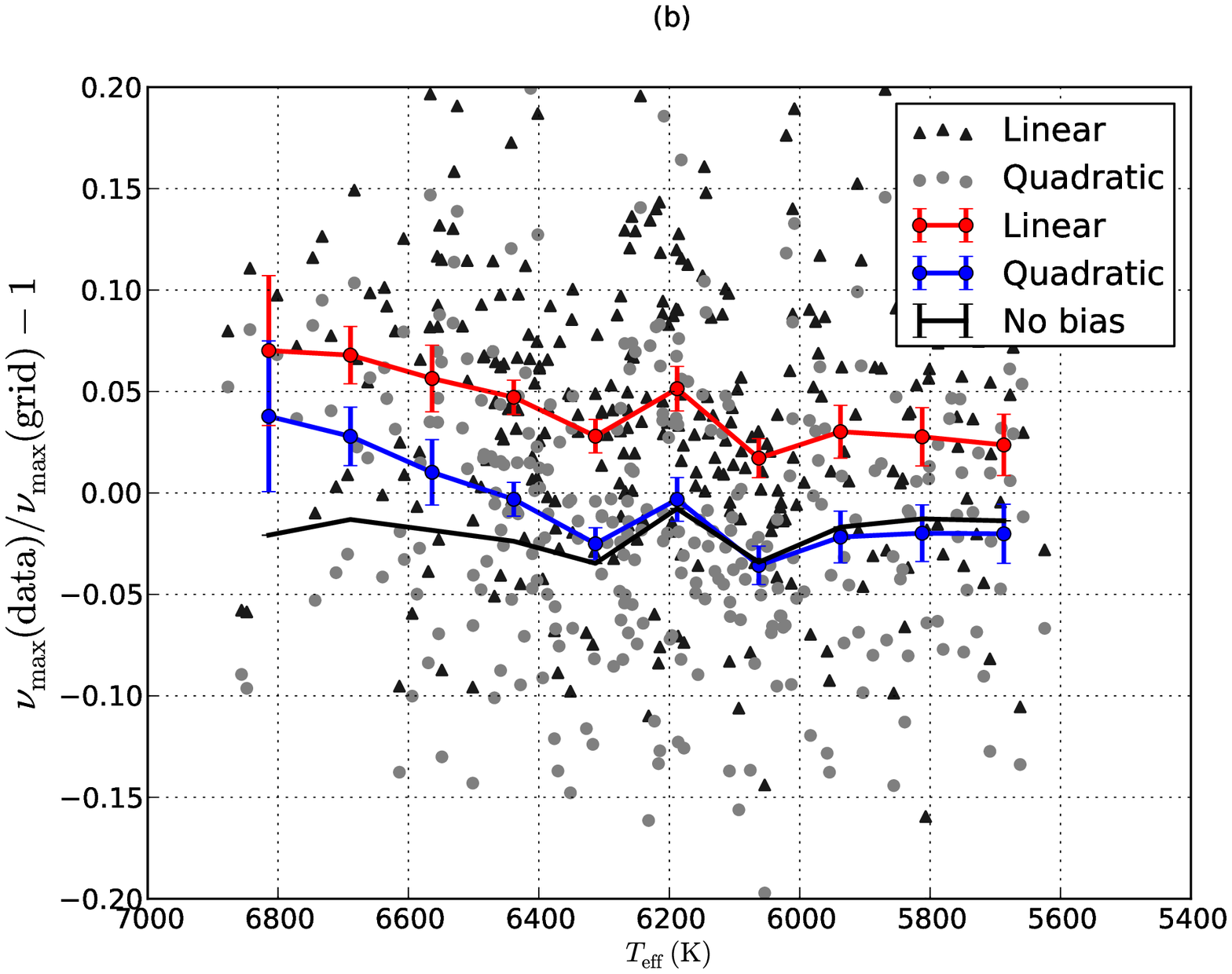}
              \epsfxsize=9.0cm\epsfbox{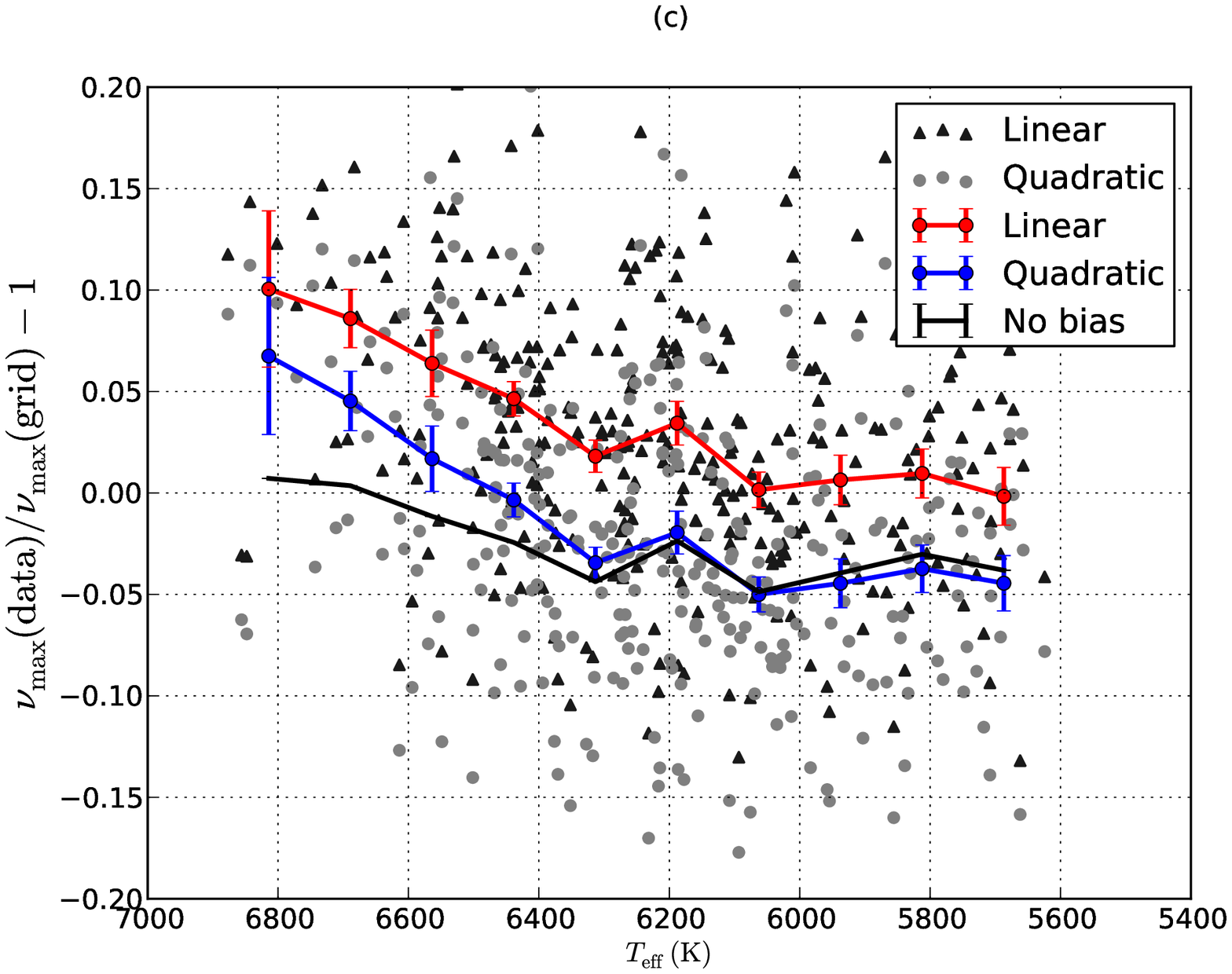}}
 \centerline {\epsfxsize=9.0cm\epsfbox{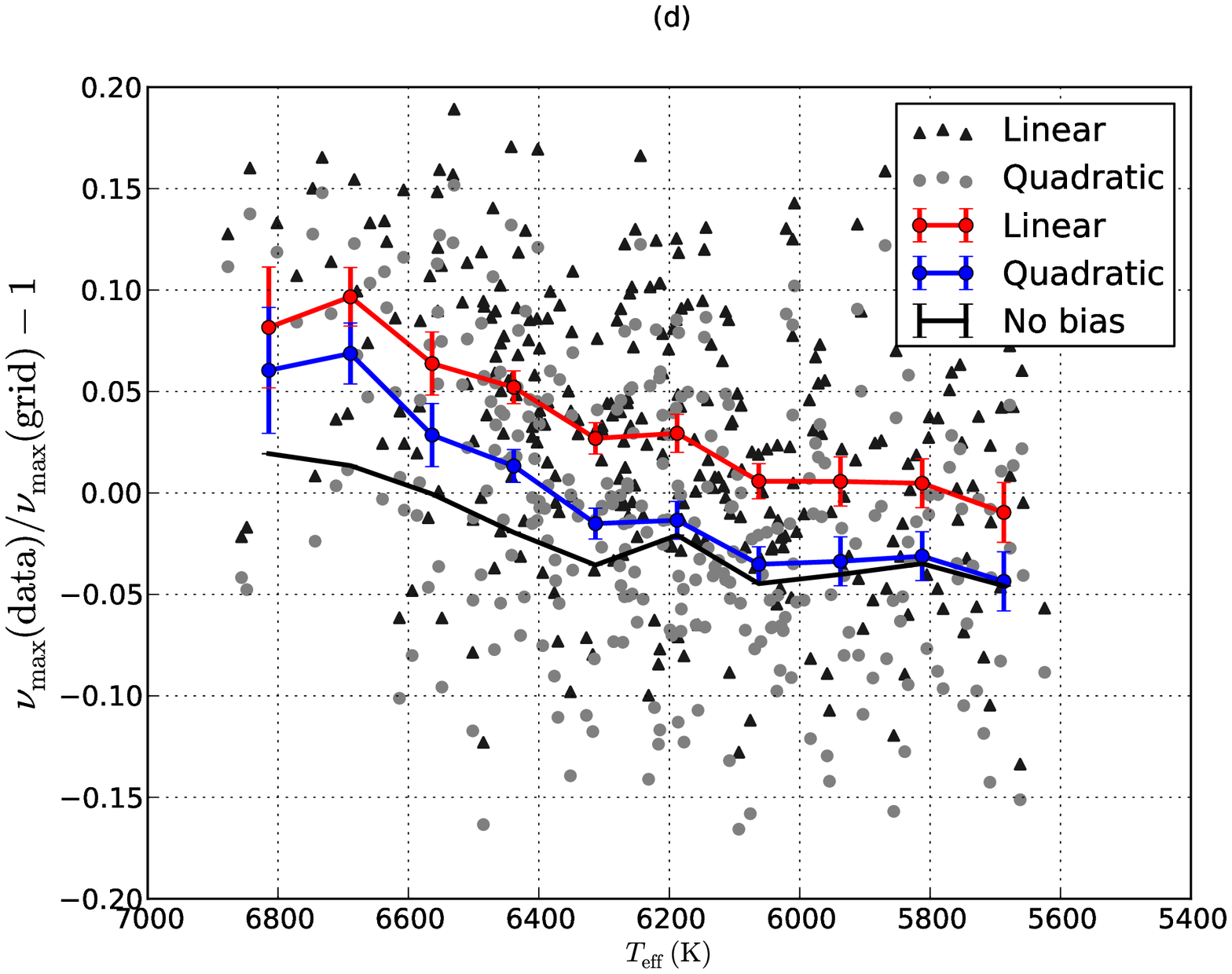}
              \epsfxsize=9.0cm\epsfbox{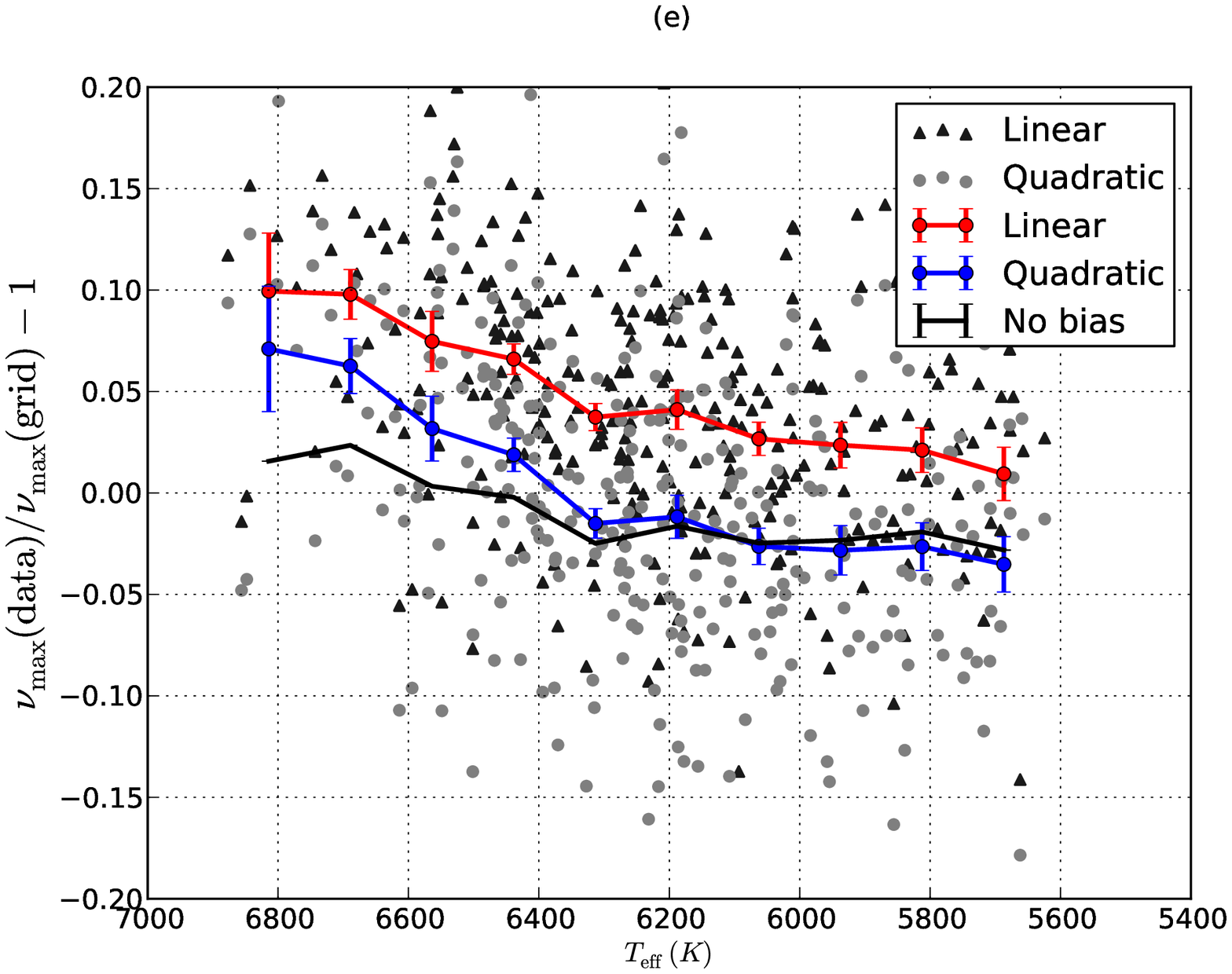}}

 \caption{Results from all grid pipelines, using an input [Fe/H] of
   $-0.05\,\rm dex$, for artificial datasets having a $T_{\rm
     eff}$-dependent bias imposed on the solar-calibrated $\nu_{\rm
     max}$ scaling relation. Panel (a): Linear and quadratic biases,
   ${\cal F}(T_{\rm eff})$ (Equation~\ref{eq:off1}), imposed on the
   artificial data. Other panels: results of testing the biased
   artificial data. Plotted are the fractional differences $\nu_{\rm
     max}({\rm data})/\nu_{\rm max}({\rm grid}) - 1$ for each biased
   set, returned by: BeSPP in frequency mode (b); BeSPP run in scaling
   mode (c); YB ALL (d); and PARAM (e). Each panel also shows for
   comparison the result from the unbiased data (see
   Figs.~\ref{fig:art1} and~\ref{fig:art2}).}

 \label{fig:art3}
\end{figure*}


Fig.~\ref{fig:art3} presents results for the artificial sets which
have known biases in $\nu_{\rm max}$.  Panel (a) shows the linear and
quadratic biases, ${\cal F}(T_{\rm eff})$ (Equation~\ref{eq:off1}),
which we imposed on the artificial data. The other panels show results
of testing the biased artificial data, with fractional differences
plotted for each biased set (see figure legends). Also plotted for
reference (black lines) are the recovered trends from the unbiased
data (see Figs.~\ref{fig:art1} and~\ref{fig:art2}). The panels show
returned fractional differences from: BeSPP in frequency mode (b);
BeSPP run in scaling mode (c); YB ALL (d); and PARAM (e).

The results are again encouraging. Evidence of the injected bias is
clearly present in the results. We also see the offset between the two
bias trends, although we would not be able to tell the difference
between the shapes of the trends.  Nevertheless, we are able to
conclude that for the level of bias tested here, it would be possible
to discriminate between the no-bias and bias cases.

 \subsection{Results from \emph{Kepler} data}
 \label{sec:realres}


\begin{figure*}

 \centerline{\epsfxsize=9.0cm\epsfbox{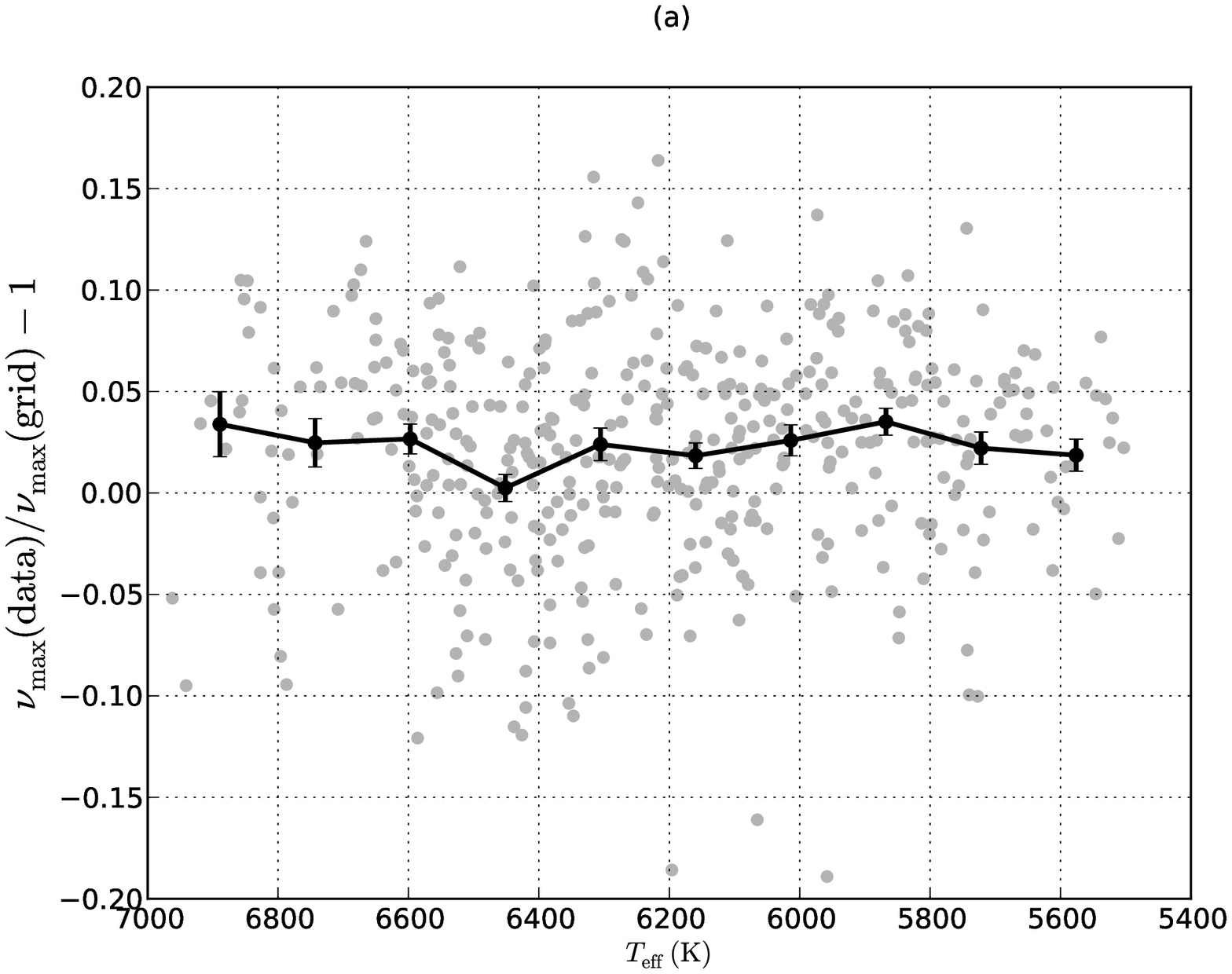}
             \epsfxsize=9.0cm\epsfbox{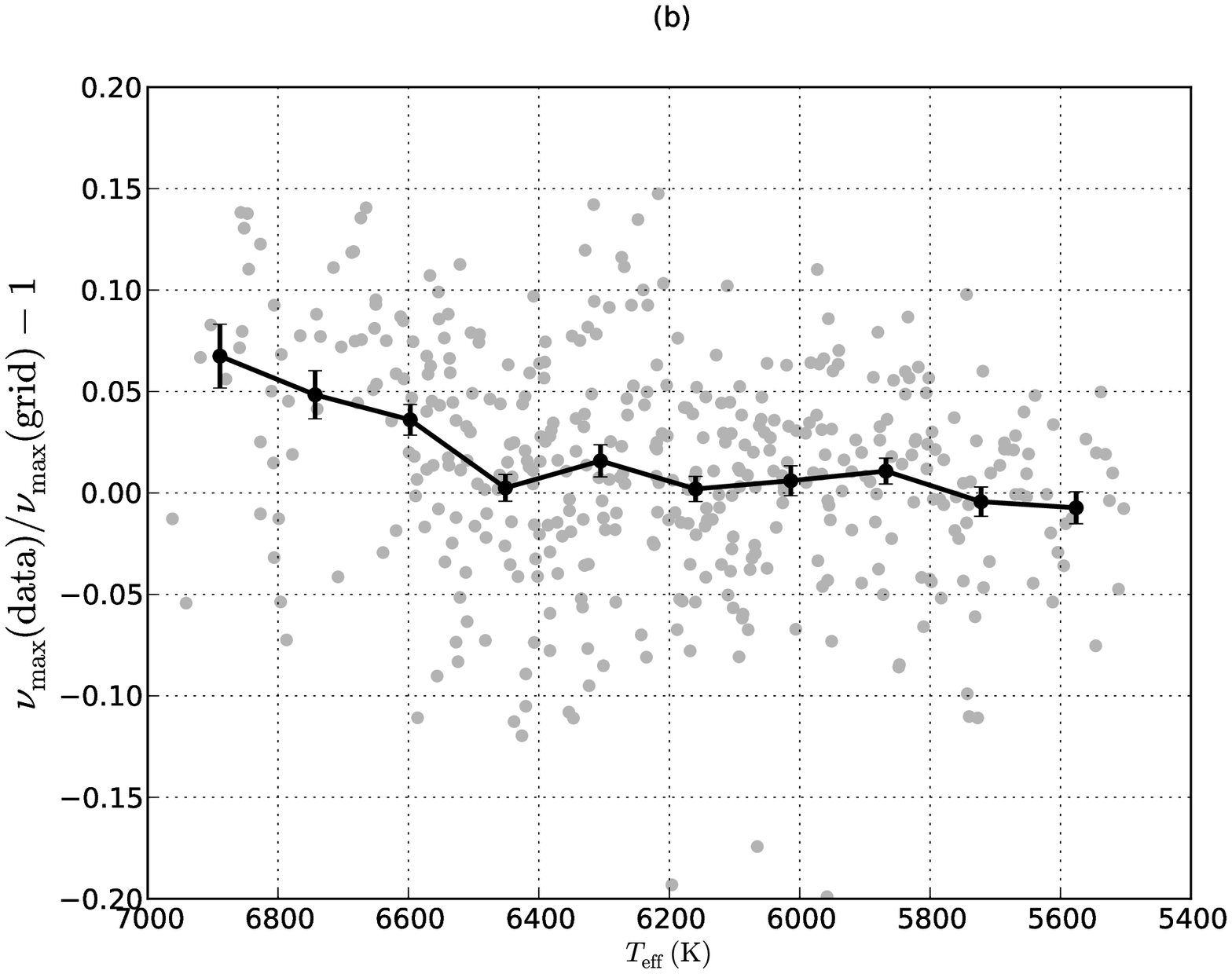}}
 \centerline{\epsfxsize=9.0cm\epsfbox{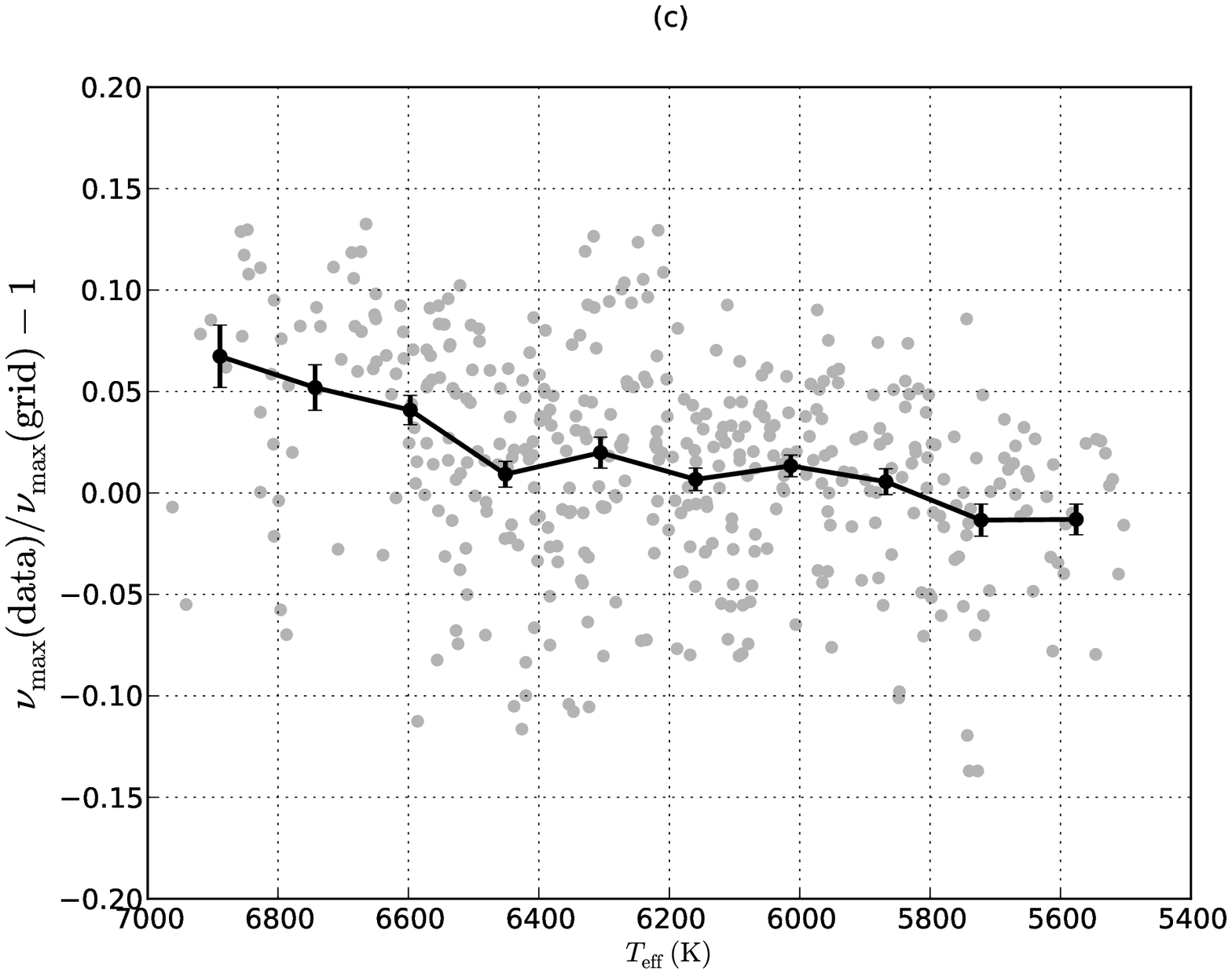}
             \epsfxsize=9.0cm\epsfbox{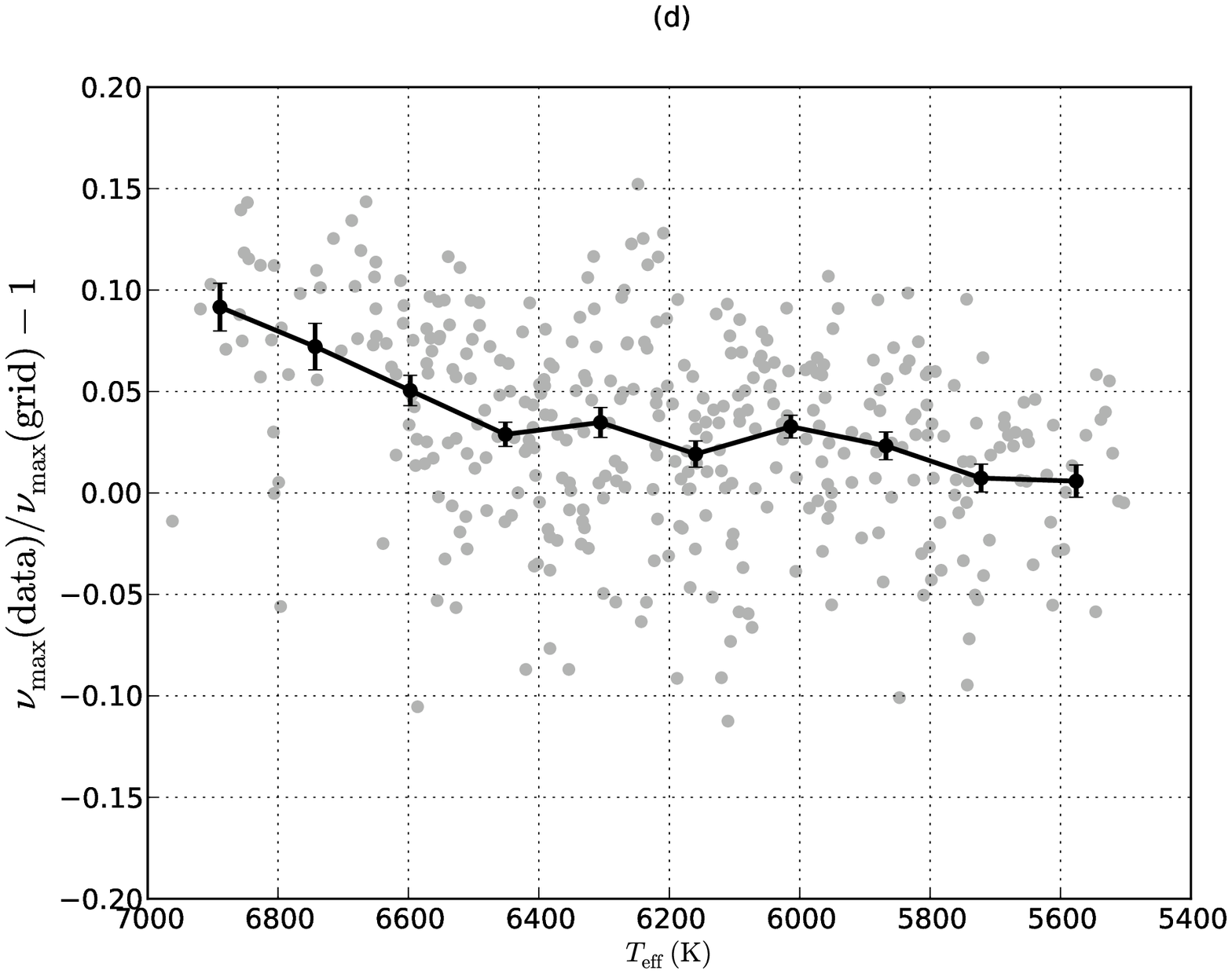}}
 \centerline{\epsfxsize=9.0cm\epsfbox{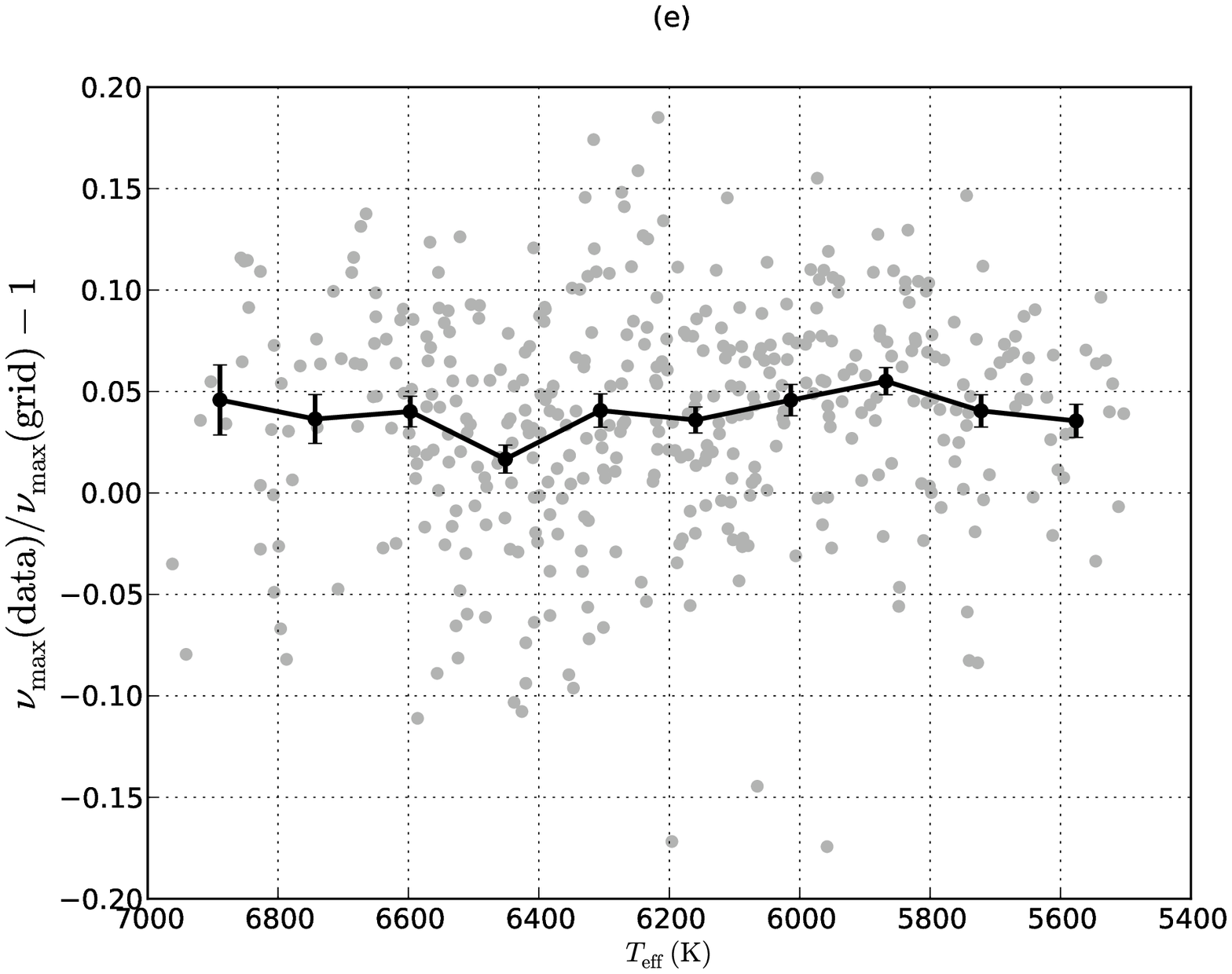}}

 \caption{Results from analysis of the real \emph{Kepler} sample of
   solar-type stars, showing the fractional differences $\nu_{\rm
     max}({\rm data})/\nu_{\rm max}({\rm grid}) - 1$ returned by the
   different pipelines. The top four panels show results with ${\rm
     [Fe/H]}=-0.05\,\rm dex$ used as input, as returned by: BeSPP in
   frequency mode (a); BeSPP run in scaling mode (b); YB ALL (c); and
   PARAM (d). Panel (e): BeSPP frequency-mode results form using an
   input [Fe/H] of $-0.20\,\rm dex$ for every star.}

 \label{fig:real1}
\end{figure*}


Fig.~\ref{fig:real1} shows results from analysing the selected sample
of \emph{Kepler} solar-type stars, which all have complementary
photometric data. The fractional differences $\nu_{\rm max}({\rm
  data})/\nu_{\rm max}({\rm grid})-1$ are plotted for the different
pipelines, including results obtained using BeSPP in both modes of
operation. The top four panels show results with ${\rm
  [Fe/H]}=-0.05\,\rm dex$ used as input, as returned by: BeSPP in
frequency mode (a); BeSPP run in scaling mode (b); YB ALL (c); and
PARAM (d).

The trends observed here bear a striking resemblance to those given by
the results on the artificial data that follow the classic $\nu_{\rm
  max}$ scaling relation. The trend in the BeSPP frequency-mode
\emph{Kepler} results on the real data is flat in $T_{\rm eff}$, and
consistent with a $gT_{\rm eff}^{-1/2}$ like scaling at the level of
precision of the data.

\begin{figure*}

 \centerline {\epsfxsize=9.0cm\epsfbox{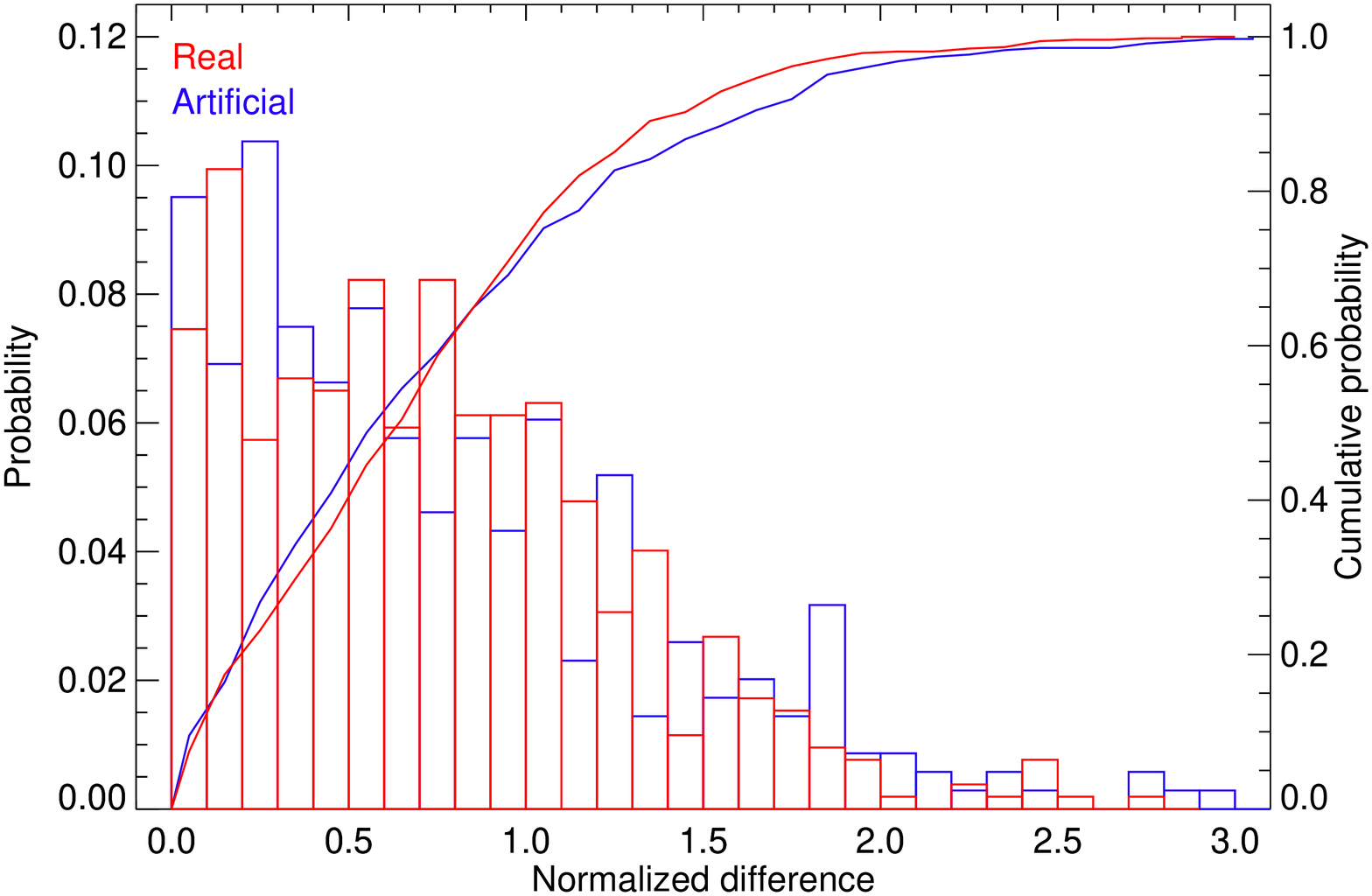}}

 \caption{Bar plots (associated to left-hand ordinate scale):
   histograms of the normalized distribution of fractional differences
   $\nu_{\rm max}({\rm data}) /\nu_{\rm max}({\rm grid})-1$ for the
   artificial no-bias (blue) and real (red) BeSPP frequency-mode
   results, which were plotted in Figs.~\ref{fig:art1}
   and~\ref{fig:real1} respectively.  Each fractional difference was
   subtracted from the respective mean trend-line of its sample, and
   then normalized by its formal uncertainty, that uncertainty having
   been propagated from the individual formal uncertainties on
   $\nu_{\rm max}({\rm data})$ and $\nu_{\rm max}({\rm grid})$. Lines
   (associated to right-hand ordinate scale): the cumulative
   distributions of the histogram data.}

 \label{fig:errs}
\end{figure*}

The scatter in the fractional differences of the real data in
Fig.~\ref{fig:real1} is entirely consistent with statistical scatter,
given the formal uncertainties. The histograms in Fig.~\ref{fig:errs}
show the normalized distributions of fractional differences for the
real (red) and artificial no-bias (blue) BeSPP frequency-mode results.
Here, each $\nu_{\rm max}({\rm data}) /\nu_{\rm max}({\rm grid})-1$
was subtracted from the respective mean trend-line of its sample, and
then normalized by its formal uncertainty, that uncertainty having
been propagated from the individual formal uncertainties on $\nu_{\rm
  max}({\rm data})$ and $\nu_{\rm max}({\rm grid})$. The lines show
the cumulative distributions of the histogram data. The real and
artificial data histograms follow one another very closely.


We may quantify further the adherence of the BeSPP frequency-mode
results to a $gT_{\rm eff}^{-1/2}$ like scaling, i.e., a flat
trend. If we fit a simple linear model to \emph{all} the data, the
best-fitting gradient implies a linear bias of $0.16 \pm 0.10$\,per
cent per 100\,K. We may therefore rule out departures from a $gT_{\rm
  eff}^{-1/2}$ scaling at the level of $\simeq 1.5\,\rm per cent$ over
the full $\Delta T \simeq 1560\,\rm K$ range tested in $T_{\rm eff}$.

As noted in Section~\ref{sec:intro}, the scaling relations
(Equations~\ref{eq:dnu} and~\ref{eq:numax}) may be manipulated to give
expressions for the stellar radius, $R$, and mass, $M$. The
dependencies of the resulting expressions imply that any bias in
$\nu_{\rm max}$ propagates to give bias in the inferred properties of
 \[
 \left(\frac{\delta R}{R}\right) = \left(\frac{\delta\nu_{\rm
     max}}{\nu_{\rm max}}\right),
 \]
and 
 \[
 \left(\frac{\delta M}{M}\right) = 3\left(\frac{\delta\nu_{\rm
     max}}{\nu_{\rm max}}\right).
 \]
Departures from the classic scaling of $\simeq 1.5\,\rm per cent$
across the full range (see above) therefore fix 1-$\sigma$ upper
limits on changes in any bias in inferred properties with changing
$T_{\rm eff}$ of $\delta R/R \le 1.5\,\rm per cent$ and $\delta M/M
\le 4.5\,\rm per cent$. Since the $\nu_{\rm max}$ scaling relation
implies
 \[
 \left(\frac{\delta g}{g}\right) = \left(\frac{\delta\nu_{\rm
     max}}{\nu_{\rm max}}\right),
 \]
bias in surface gravities inferred from use of the relation will be
similar to that for the radii above (which corresponds to a bias in
$\log g$ of $\simeq 0.006\,\rm dex$).

We also fitted data in overlapping $\Delta T_{\rm eff} = 500\,K$-wide
bins, moving systematically through the full range to test the
temperature dependence of any departures from the scaling. Throughout
most of the range we obtain 1-$\sigma$ uncertainties on the
best-fitting gradients of $\simeq 0.4$\,per cent per 100\,K. The
limits increase to $\simeq 0.5$\,per cent per 100\,K at temperatures
above about 6200\,K, rising to $\simeq 1$\,per cent per 100\,K in the
highest range we fitted, which started at 6450\,K.

Our results neverthless imply some uncertainty over the
\emph{absolute} calibration of the $\nu_{\rm max}$ scaling.  Panel (e)
in Fig.~\ref{fig:real1} shows the impact on the BeSPP frequency-mode
results of using a different input [Fe/H] for each star, here
$-0.20\,\rm dex$ instead of the value of $-0.05\,\rm dex$ used for the
top left-hand panel. Just like the artificial data in
Fig.~\ref{fig:art1}, we see a systemetic shift in the absolute offset,
from just over 3\,per cent for an input [Fe/H] of $-0.05\,\rm dex$ to
around 4\,per cent for an input [Fe/H] of $-0.20\,\rm dex$.  Assuming
that the overall shift results from the quadratic addition of
uncorrelated effects, this implies a systematic shift due to the
uncertainty in [Fe/H] of around 2.5\,per cent. It is worth noting that
if we analyse the smaller sample of real stars having spectroscopic
data from Bruntt et al. (2012) -- which have much more tightly
constrained input [Fe/H] -- we get the same $\approx 3$\,per cent
offset as the $-0.05\,\rm dex$ full-sample case.

Accounting for the above still leaves us with an uncertainty in the
absolute calibration of around 3\,per cent (again, assuming that the
contributions are uncorrelated). We recall from discussions in
previous sections that we would expect this to reflect differences
between the physics and structures of the real stars (including
near-surface effects) and those of the grid models.

An analysis performed on $\Delta T_{\rm eff} = 500\,K$-wide bins finds
no evidence for any significant deviation of the offset with $T_{\rm
  eff}$.  The results imply that if there are any absolute errors in
the calibration, all inferred properties will be biased by the
\emph{same} fractional amount.  For the solar-calibrated
Equation~\ref{eq:numax} the calibration is provided by the value
$\nu_{\rm max\,\odot} = 3090\,\rm \mu Hz$.  Guided by the results
above, a 3\,per cent uncertainty in the overall calibration translates
to an error in the the calibrating frequency of around $100\,\rm \mu
Hz$.

 \section{Conclusion}
 \label{sec:conc}

Asteroseismic scaling relations for solar-like oscillators are
becoming increasingly important as they receive more widespread use in
building large catalogues of stellar properties.

We have tested the scaling relation for the global asteroseismic
parameter $\nu_{\rm max}$, the frequency at which a solar-like
oscillator presents its strongest observed pulsation amplitude.  The
classic scaling relation assumes that $\nu_{\rm max}$ scales with
surface gravity and effective temperature according to $gT_{\rm
  eff}^{-1/2}$.  We have tested how well the detected oscillations in
a large sample of solar-type stars observed by the NASA \emph{Kepler}
Mission adhere to this relation by comparing the observed $\nu_{\rm
  max}$ of the stars with independent estimates of the combination
$g\,T_{\rm eff}^{-1/2}$.

Our results rule out departures from the classic $\nu_{\rm max}$
scaling at the level of $\simeq 1.5\,\rm per cent$ over the full
$\simeq 1560\,\rm K$ range in $T_{\rm eff}$ that we tested.  There is
some uncertainty over the absolute calibration of the scaling, but any
variation with $T_{\rm eff}$ is evidently small, with limits similar
to those above.

\subsection*{ACKNOWLEDGMENTS}

The authors would like to thank L. Girardi and T. S. Rodrigues for use
of the most recent PARAM code. Funding for this Discovery mission is
provided by NASA's Science Mission Directorate. The authors wish to
thank the entire \emph{Kepler} team, without whom these results would
not be possible.  The research leading to these results has received
funding from the the UK Science and Technology Facilities Council
(STFC) and the European Community's Seventh Framework Programme
([FP7/2007-2013]) under grant agreement no. 312844 (SPACEINN).
H.R.C. acknowledges support from grant BEX 8569/12-6
(CAPES/Birmingham). Funding for the Stellar Astrophysics Centre is
provided by The Danish National Research Foundation (Grant agreement
no.: DNRF106). S.B. acknowledges partial support from NSF grant
AST-1105930 and NASA grant NNX13AE70G.  A.M.S. is partially supported
by grants ESP2013-41268-R (MINECO) and 2014SGR-1458 (Generalitat of
Catalunya).


\begin{thebibliography}{}

\bibitem{ba10} Basu, S., Chaplin, W. J., Elsworth, Y., 2010, ApJ, 710,
  1596

\bibitem{ba12} Basu, S., Verner, G. A., Chaplin, W. J., Elsworth,
  Y. 2012, ApJ, 746, 76

\bibitem{bedd11} Bedding TR. 2014. In Asteroseismology, Canary Islands
  Winter School of Astrophysics, Vol. 23, eds. P. L. Pall/'e,
  C. Esteban.  Cambridge, UK: Cambridge Univ. Press, p. 60

\bibitem{bel11} Belkacem, K., Goupil, M. J., Dupret, M. A., et al,
  2011, A\&A, 530, 142

\bibitem{bell13} Belkacem, K., Samadi, R., Mosser, B., Goupil, M.-J.,
  Ludwig, H.-G., 2013, ASPC, 479, 61

\bibitem{bre12} Bressan, A. Marigo, P., Girardi, L., Salasnich, B.,
  Dal Cero, C., Rubele, S., Nanni, A., 2012, MNRAS, 427, 127

\bibitem{} Brown, T. M., Gilliland, R. L., Noyes, R. W., Ramsey,
  L. W., 1991, ApJ, 368, 599

\bibitem{} Brown, T. M., Latham, D. W., Everett, M. E., Esquerdo,
  G. A. 2011, AJ, 142, 111

\bibitem{} Brogaard, K., VandenBerg, D. A., Bruntt, H., et al., 2012,
  A\&A, 543, A106

\bibitem{br10} Bruntt, H., Bedding, T. R., Quirion, P. O., et
  al. 2010, MNRAS, 405, 1907

\bibitem{br12} Bruntt, H., Basu, S., Smalley, B., et al. 2012, MNRAS,
  423, 122

\bibitem{car08} Carigi, L., Peimbert, M., 2008, RMXAA, 44, 341

\bibitem{cas10} Casagrande, L., Ram\'irez, I., Mel\'eendez, J., et
  al. 2010, A\&A, 512, 54

\bibitem{cas14} Casagrande, L., Silva Aguirre, V., Stello, D., et al.,
  2013, ApJ, 787, 110

\bibitem{} Chaplin, W. J., Houdek, G., Appourchaux, T., et al. 2008,
A\&A, 485, 813

\bibitem{wjc13b} Chaplin, W. J., Miglio, A., 2013, ARAA, 51, 353

\bibitem{chaplin14} Chaplin, W. J., Basu, S., Huber, D., et al., 2014,
  ApJS, 210, 1

\bibitem{chi82} Chiosi, C., Matteucci, F. M., 1982, A\&A, 105, 140

\bibitem{chr93} Christensen-Dalsgaard, J. 1993, Proc. GONG 1992,
  Seismic Investigation of the Sun and Stars (ASP Conf. Ser. 42),
  ed. T. M. Brown (San Francisco, CA: ASP), p. 347

\bibitem{cy03} Cyburt, R. H., Fields, B. D., Olive, K. A.,
  Phys. Lett. B., 567, 227

\bibitem[]{dsil06} da Silva L., et al., 2006, A\&A, 458, 609

\bibitem{pd04} Demarque, P., Woo, J.-H., Kim, Y.-C., Yi, S. K. 2004,
  ApJS, 155, 667

\bibitem{do89} Dotter, A., et al. 2008, ApJS, 178, 89

\bibitem{gai11} Gai, N., Basu, S., Chaplin, W. J., Elsworth, Y., 2011,
  ApJ, 730, 63


\bibitem{grev93} Grevesse, N., Noels, A., 1993, PhST, 47, 133

\bibitem{huber12} Huber, D., Ireland, M. J., Bedding, T. R., et al.,
  2012, ApJ, 760, 32

\bibitem{} Huber, D., Chaplin, W. J., Christensen-Dalsgaard, J., et
  al., 2013, ApJ, 767, 127

\bibitem{} Kjeldsen, H., Bedding, T. R., 1995, A\&A, 293, 87


\bibitem{mig12a} Miglio, A., 2012, in: Red Giants as Probes of the
  Structure and Evolution of the Milky Way.  Berlin: Springer, Miglio,
  A., Montalb\'an J., Noels, A., eds., p.~11

\bibitem{mig12b} Miglio, A., Brogaard, K., Stello, D., et al., 2012,
  MNRAS, 419, 2077

\bibitem{mig13} Miglio, A., Chiappini, C., Morel, T., et al., 2013,
  MNRAS, 429, 423

\bibitem{mo14} Morel, T., Miglio, A., Lagarde, N., et al., 2014, A\&A,
  564, 119

\bibitem{} Mosser, B., Michel, E., Belkacem, K., et al., 2013, A\&A,
  550, 126

\bibitem{pax11} Paxton, B., Bildsten, L., Dotter, A., et al., 2011,
  ApJS, 192, 3

\bibitem{piet04} Pietrinferni, A., Cassisi, S., Salaris, M., \&
  Castelli, F. 2004, ApJ, 612, 168

 \bibitem{pin12} Pinsonneault, M. H., An, D., Molenda-\.Zakowicz, J.,
   et al. 2012, ApJS, 199, 30

\bibitem{pin14} Pinsonneault, M. H., Elsworth, Y., Epstein, C., et
  al. 2014, ApJS, 215, 19

\bibitem{ree12} Reese, D. R., Marques, J. P., Goupil, M. J., Thompson,
  M. J., Deheuvels, S., 2012, A\&A, 539, 63

\bibitem{rod14} Rodrigues, T. S., Girardi, L., Miglio, A., et al.,
  2014, MNRAS, 445, 2758

\bibitem{sand13} Sandquist, E. L., Mathieu, R. D., Brogaard, K., et
  al., 2013, ApJ, 762, 58

\bibitem{aldo13} Serenelli, A. M., Bergemann, M., Ruchti, G.,
  Casagrande, L., 2013, MNRAS, 429, 3645

\bibitem{vsa11} Silva Aguirre, V., Chaplin, W. J., Ballot, J., et
  al. 2011, ApJL, 740, L2

\bibitem{vsa12} Silva Aguirre, V., Casagrande, L., Basu, S., et al. 2012,
  ApJ, 757, 99

\bibitem{vsa13} Silva Aguirre, V., Basu, S., Brand\~ao, I. M., et
  al. 2013, ApJ, 769, 141

\bibitem{sk06} Skrutskie, M. F., Cutri, R. M., Stiening, R., et
  al. 2006, AJ, 131, 1163

\bibitem{tas80} Tassoul, M., 1980, ApJS, 43, 469

\bibitem{town13} Townsend, R. H. D., Teitler, S. A., 2013, MNRAS, 435,
  3406

\bibitem{ulr86} Ulrich R. K., 1986, ApJ, 306, L37

\bibitem{white11} White, T. R., Bedding, T. R., Stello, D.,
  Christensen-Dalsgaard, J., Huber, D., Kjeldsen, H., 2011, ApJ, 743,
  161

\bibitem{white13} White, T. R., Huber, D., Maestro, V., et al., 2013,
  MNRAS, 433, 1262


\end{thebibliography}
\end{document}